\documentclass[journal,12pt,onecolumn,draftclsnofoot,]{IEEEtran}

\usepackage[T1]{fontenc}
\usepackage{amsmath}
\usepackage[cmintegrals]{newtxmath}
\usepackage{graphicx}
\usepackage{url}
\usepackage{cite}
\usepackage{algorithm}
\usepackage{algpseudocode}

\makeatletter
\def\BState{\State\hskip-\ALG@thistlm}
\makeatother

\usepackage{epstopdf}
\usepackage{amsmath}
\usepackage{amsxtra}
\usepackage{amstext}
\usepackage{amssymb}
\usepackage{latexsym}
\usepackage{dsfont} 
\usepackage{color}
\usepackage{multirow}
\usepackage{float}
\usepackage{hyperref}

\usepackage{units}
\usepackage{cases}
\usepackage[ subrefformat=parens,labelformat=parens, caption=false, font=footnotesize]{subfig}
\usepackage{mathtools}
\mathtoolsset{showonlyrefs} 

\newcommand{\mbf}[1]{\mathbf{#1}}

\newcommand{\nth}[1]{{#1}{\text{th}}}

\newcommand{\abs}[1]{\left|{#1}\right|}
\newcommand{\norm}[1]{\left\|{#1}\right\|}

\DeclareMathOperator*{\argmin}{arg\,min}

\usepackage{mathtools}

\newcommand{\ML}{\mathrm{ML}}
\newcommand{\PML}{\mathrm{PML}}
\newcommand{\NC}{\mathrm{N/C}}
\newcommand{\PNC}{\mathrm{PN/C}}
\newcommand{\CD}{\mathrm{CD}}
\newcommand{\PCD}{\mathrm{PCD}}

\newcommand{\LORD}{\mathrm{LORD}}
\newcommand{\SSD}{\mathrm{SSD}}
\newcommand{\VSSD}{\mathrm{VSSD}}
\newcommand{\SLORD}{\mathrm{SLORD}}
\newcommand{\SSSD}{\mathrm{SSSD}}

\newcommand{\RML}{\mathrm{RML}}
\newcommand{\RAD}{\mathrm{RAD}}

\newcommand{\Prb}{\mathrm{Pr}}
\newcommand{\err}{\mathrm{err}}
\newcommand{\A}{\mathrm{A}}
\newcommand{\B}{\mathrm{B}}
\newcommand{\C}{\mathrm{C}}
\newcommand{\D}{\mathrm{D}}

\newcommand{\F}{\mathrm{F}}

\newcommand{\Tr}{\mathrm{Tr}}
\newcommand{\Rp}{\mathring{\mbf{R}}}
\newcommand{\rp}{\mathring{r}}
\newcommand{\rrp}{\mathring{\mbf{r}}}

\begin{document}
\bstctlcite{IEEEexample_new:BSTcontrol}

\title{Large MIMO Detection Schemes Based on Channel Puncturing: Performance and Complexity Analysis}

\author{Hadi~Sarieddeen,
        Mohammad~M.~Mansour,
        and~Ali~Chehab
\thanks{
H. Sarieddeen, M. M. Mansour, and A. Chehab are with the Department of Electrical
and Computer Engineering, American University of Beirut, Beirut 1107 2020,
Lebanon (e-mail: has63@aub.edu.lb; mmansour@aub.edu.lb; chehab@aub.edu.lb). 
}
}

\maketitle
\vspace{-0.1in}
\begin{abstract}
A family of low-complexity detection schemes based on channel matrix puncturing targeted for large multiple-input multiple-output (MIMO) systems is proposed. It is well-known that the computational cost of MIMO detection based on QR decomposition is directly proportional to the number of non-zero entries involved in back-substitution and slicing operations in the triangularized channel matrix, which can be too high for low-latency applications involving large MIMO dimensions. By systematically puncturing the channel to have a specific structure, it is demonstrated that the detection process can be accelerated by employing standard schemes such as chase detection, list detection, nulling-and-cancellation detection, and sub-space detection on the transformed matrix. The performance of these schemes is characterized and analyzed mathematically, and bounds on the achievable diversity gain and probability of bit error are derived. Surprisingly, it is shown that puncturing does not negatively impact the receive diversity gain in hard-output detectors. The analysis is extended to soft-output detection when computing per-layer bit log-likelihood ratios; it is shown that significant performance gains are attainable by ordering the layer of interest to be at the root when puncturing the channel. Simulations of coded and uncoded scenarios certify that the proposed schemes scale up efficiently both in the number of antennas and constellation size, as well as in the presence of correlated channels. In particular, soft-output per-layer sub-space detection is shown to achieve a $\unit[2.5]{dB}$ $\mathsf{SNR}$ gain at $10^{-4}$ bit error rate in $256$-QAM $16\!\times\!16$ MIMO, while saving $77\%$ of nulling-and-cancellation computations.
\end{abstract}
\vspace{-0.125in}
\begin{IEEEkeywords}\vspace{-0.125in}
Channel puncturing, large MIMO, chase detector, sub-space detector, BER performance
\end{IEEEkeywords}

\section{Introduction}
\IEEEPARstart{M}IMO technology is a technique that exploits the spatial dimension by adding more antennas \cite{2003_Paulraj} to increase spectral efficiency and network capacity. However, conventional MIMO configurations fall short of providing the required spatial diversity in the upcoming fifth generation (5G) mobile communication standard, which promises to connect billions of devices and achieve several gigabit-per-second data rates. Towards this end, \emph{massive} MIMO has been introduced \cite{Larsson_2014}, in which few hundred antennas serve tens of terminals over time and frequency resources.

Despite the extensive work on massive MIMO, \emph{large} MIMO will also play an important role in the future. Large MIMO systems use tens of antennas in communication terminals, and can afford large number of antennas on both the transmitter and the receiver sides \cite{chockalingam2014large}, such as for example $8\times8$, $16\times16$, $32\times32$, and $64\times64$ configurations. Large point-to-point MIMO wireless links are of specific interest in 5G for high-speed wireless backhaul connectivity between base stations (BSs). Also, multipoint-to-point large multiuser MIMO can be used in 5G in the uplink when the number of served transmitting users is less than, but comparable to, the number of BS antennas. Nevertheless, large MIMO can also be considered for point-to-multipoint downlink multiuser MIMO (MU-MIMO)~\cite{2011_Duplicy}, whether in enhanced versions of the current wireless communications standards, or in 5G, where users sharing the same physical resource blocks are chosen based on the degree of orthogonality of their cascaded precoder and channel.

After being traditionally driven by diversity-multiplexing tradeoffs, recent wireless communication system designs have been driven by two factors; system performance in terms of throughput and bit error rate, and system complexity in terms of processing latency and computational complexity. The performance of MIMO systems is largely determined by the detection scheme at the receiver side; various schemes provide different performance-complexity tradeoffs \cite{bai2014low}. Linear detectors, such as zero forcing (ZF) and minimum mean square error (MMSE), are the least-complex, but the least-optimal as well. On the other hand, maximum likelihood (ML) detectors are optimal but most computationally intensive, with complexity that grows exponentially with the number of antennas. Several sub-optimal detectors fill the spectrum in between, including sphere decoders (SD) and their variants \cite{Viterbo,barbero2008fixing,2014_sphereP1_mansour,2014_sphereP2_mansour}. Moreover, in addition to conventional hard-output (HO) detectors, soft-output (SO) detectors play an important role in near-capacity achieving systems, but are more complex because they require processing significantly more signal combinations to generate reliability information.

In massive MIMO systems, linear detectors achieve near-optimal performance by exploiting the channel hardening effect \cite{Ngo_2013}, and approximate matrix inversions via Neumann series approximations \cite{Rosario_2016} are used for practical implementations. However, large MIMO systems do not have very large receive-to-transmit antenna ratios. Hence, they cannot achieve the performance gains of asymmetric massive MIMO systems, and they do not allow for similar practical implementations, where Neumann series expansions fail to converge. For large MIMO systems, the detection schemes in the literature are grouped into several areas: detection based on local search \cite{Vardhan_2008,Li_2010}; detection based on meta-heuristics \cite{Datta_2010,Srinidhi_2011}; detection via message passing on graphical models \cite{Goldberger_2011,Narasimhan_2014}; lattice reduction (LR) aided detection \cite{Wubben_2011,Zhou_LR_2013}; and detection using Monte Carlo sampling \cite{Datta_2013}. However, for these schemes to achieve a near-ML performance with high orders of antennas and modulation constellations, the entailed complexity would be prohibitive.

A popular family of MIMO detectors that achieves good performance-complexity tradeoffs employs non-linear subset-stream detection. The nulling-and-cancellation (N/C) detector \cite{choi2006nulling} is a low-complexity member of this family; it consists of linear nulling followed by successive interference cancellation (SIC). The chase detector (CD) \cite{waters2008chase} is a more complex member of this family; it first creates a list of candidate decision vectors, and then chooses the best candidate from this list as a final decision. Chase detection is considered a special case of list detection. However, it differs from list sphere decoding (LSD) \cite{hochwald2003achieving} for example in the way the list is generated and administered; in LSD, list admission is based on proximity to an initial solution, while in CD, list generation is deterministic, and is done by spanning all possible sub-tree symbols emanating from the root symbol in a specific layer of interest. Furthermore, other popular subset-stream detectors exist (e.g.,~\cite{jiang2005joint,jiang2005uniform,Ariyavisitakul}), that decompose the channel matrix into lower order sub-channels to reduce the number of jointly detected streams.

All aforementioned subset-stream detectors make use of QR decomposition (QRD). However, the SO sub-space detector (SSD) \cite{ojard2008method}, transforms the channel matrix via a punctured QRD, which we refer to in this paper as WR decomposition (WRD). In~\cite{2014_mansour_SPL_WLD,2014_mansour_eurasip_WLD,Sarieddeen_WCNC_Subspace}, WRD-based SSD is generalized to allow for joint detection of arbitrary-sized subsets of decoupled streams, and efficient implementation methods are presented. The QRD-based version of this detector is called the layered orthogonal lattice detector (LORD) \cite{Siti-1,tomasoni2012hardware}, and both are special cases of CD. To the best of our knowledge, the use of punctured QRD in MIMO detectors has not been studied analytically in the literature, and its applicability to large MIMO systems has not been addressed. \\

The contributions of this paper are summarized as follows:
\begin{enumerate}
  \item We present a family of WRD-based detectors that build on popular QRD-based detectors. In particular, we propose a punctured ML (PML) detector, a punctured N/C (PN/C) detector, a punctured CD (PCD), as well as a hard-output sub-space detector.
  \item We analyze mathematically the bit error rate (BER) performance of the proposed HO detectors. First, the diversity gain is characterized and used to show that channel matrix puncturing does not negatively affect the diversity gain in HO detection. Second, the performance of these detectors is studied via a probabilistic BER characterization.
  \item We extend the study for several variations of SO detection schemes, and show that significant performance gains can be achieved with channel puncturing.
  \item We propose efficient architectures and analyze the computational complexity of the proposed detectors. We show that the computational savings are much more pronounced with large MIMO dimensions.
  \item We study the performance of the proposed detectors in the context of large MIMO with high order modulations, and in the presence of spatial channel correlation. We show that the performance of these schemes scales up efficiently with high orders, and that they are superior to their QRD-based counterparts in the presence of channel correlation.
\end{enumerate}

The remainder of the paper is organized as follows. The system model and basic reference detectors are presented in Sec.~\ref{sec:sysmodel}. The proposed WRD-based ML detector, N/C detector, CD, and SSD detection algorithms are presented in Sec.~\ref{sec:proposed}. The achievable diversity gains of these detectors are derived in Sec.~\ref{sec:diversity}, followed by a probabilistic BER characterization that describes the behaviour of the proposed approaches in Sec.~\ref{sec:probBER}. The SO versions of the detectors are then proposed in Sec.~\ref{sec:soft_output}, and an efficient architecture is proposed in Sec.~\ref{sec:complexity} alongside a complexity study. Finally, simulation results are presented in Sec.~\ref{sec:results}.

Regarding notation, bold upper case, bold lower case, and lower case letters correspond to matrices, vectors, and scalars, respectively. Scalar norms, vector $\text{L}_2$ norms, and Frobenius norms are denoted by $\abs{\cdot}$, $\norm{\cdot}$, and $\norm{\cdot}_{\F}$, respectively. $\mathsf{E}[\cdot]$, $\Tr(\cdot)$, $\Re(\cdot)$, $(\cdot)^{T}$, and $(\cdot)^{*}$, stand for the expected value, trace function, real part, transpose, and conjugate transpose, respectively. $\mathcal{N}(\cdot)$ refers to normal distribution, and $Q(\cdot)$ refers to the Q-function, where $Q(x)\!=\!\int_x^{\infty} e^{-z^2/2}/\sqrt{2\pi}\ dz$. $\Rp\!=\![\rp_{ij}]$ is a punctured matrix $\mbf{R}$ with entries $\rp_{ij}$, and $\mbf{I}_N$ is an identity matrix of size $N$. Detector ML optimality is in the log-max sense.

%
\section{System Model and Reference Detectors}\label{sec:sysmodel}

%
\subsection{System Model}\label{sec:model}
We consider spatial multiplexing in a MIMO system with $N$ transmit antennas and $M = N$ receive antennas. The equivalent complex baseband input-output system relation is given by
\begin{equation}\label{eq:sysmodel}
  \mbf{y} = \mbf{H}\mbf{x} + \mbf{n},
\end{equation}
where $\mbf{y}\in\mathcal{C}^{M\times1}$ is the received complex vector, $\mbf{H}\in\mathcal{C}^{M\times N}$ is the channel matrix with entries that are assumed to be i.i.d. complex, circularly symmetric Gaussian random variables, $\mbf{x}=[x_{1}\cdots x_{n}\cdots x_{N}^{}]^{T}\in\mathcal{C}^{N\times1}$ is the transmitted symbol vector, and $\mbf{n}\in\mathcal{C}^{M\times1}$ is a complex-valued circular-symmetric Gaussian random vector with zero mean and variance $\sigma^{2}$ \big($\mathsf{E}[\mbf{n}\mbf{n}^{*}]=\sigma^{2}\mbf{I}_M$\big). Each symbol $x_{n}$, $n\in\{1,\cdots,N\}$, belongs to a normalized complex constellation $\mathcal{X}_{n}$ ($\mathsf{E}[x_{n}^{*}x_{n}^{}]\!=\!1$), and we have $\mbf{x}\!\in\! \mathcal{X} \subset \mathcal{C}^{N\times1}$, where $\mathcal{X}$ is the finite set of points on a $N$-dimensional lattice generated by all possible symbol vectors. For simplicity, we assume a uniform modulation constellation $\mathcal{M}$ on all layers, and hence $\mathcal{X}=\mathcal{M}^{N}$. The coded bit-representation of a symbol $x_n$ is denoted by $\mbf{b}_{n}\!=\!(b_{n,1},\cdots,b_{n,j},\cdots,b_{n,q})$, where $q\!=\!\lceil{\log_2(\abs{\mathcal{M}})\rceil}$ and $b_{n,j} \in \{0,1\}$ for $j=1,\cdots,q$. The signal to noise ratio ($\mathsf{SNR}$) is defined in terms of the noise variance as $\mathsf{SNR}\!=\!N/\sigma^{2}$.

At the receiver side, and assuming perfect knowledge of the channel, QRD decomposes $\mbf{H}$ as $\mbf{H}\!=\!\mbf{Q}\mbf{R}$, where $\mbf{Q}\!=\![\mbf{q}_{1}\cdots\mbf{q}_{n}\cdots \mbf{q}_{N}^{}]\!\in\!\mathcal{C}^{M \times N}$ has orthonormal columns $\mbf{q}_n\in\mathcal{C}^{M\times 1}$ and $\mbf{Q}^*\mbf{Q}\!=\!\mbf{I}_N$, and $\mbf{R}\!=\![r_{ij}^{}]\!\in\!\mathcal{C}^{N\times N}$ is a square upper-triangular matrix (UTM) with real and positive diagonal entries. The transformed receive symbol vector can then be equivalently expressed as
\begin{equation}\label{eq:sysmodel2}
  \mbf{\tilde{y}} = \mbf{Q}^{*}\mbf{y} = \mbf{R}\mbf{x} + \mbf{Q}^{*}\mbf{n},
\end{equation}
where $\mbf{Q}^{*}\mbf{n}$ and $\mbf{n}$ are statistically identical since $\mbf{Q}$ is orthonormal.

%
An ``exhaustive'' log-max ML detector searches the complete lattice $\mathcal{X}$, computing $\abs{\mathcal{M}}^N$ Euclidean distance metrics, to solve for
\begin{equation}\label{eq:ML_dist}
  \hat{\mbf{x}}^{\ML} = \min_{\mbf{x} \in \mathcal{X}}\norm{\mbf{y} - \mbf{H}\mbf{x}}^{2} = \min_{\mbf{x} \in \mathcal{X}}\norm{\mbf{\tilde{y}} - \mbf{R}\mbf{x}}^{2}.
\end{equation}
Note that the SD achieves exact log-max ML performance with less computations, by executing a tree-based search on a subset of $\mathcal{X}$, skipping vectors in the space whose partial distance already exceeds the current best distance.

%
\subsection{Nulling-and-Cancellation (N/C) Detector}\label{sec:SIC}
The N/C detector~\cite{choi2006nulling} is used in the widely known vertical Bell Labs layered space time (V-BLAST) architecture \cite{wolniansky1998v}. When combined with QRD, N/C becomes a computationally-efficient procedure which is highly sensitive to layer ordering. Nulling is performed by linearly pre-multiplying the received vector with $\mbf{Q}^{*}$, which suppresses the interference from $x_l$, $l >n$, at the $\nth{n}$ layer. This is followed by SIC (back-substitution and slicing) to suppress co-antenna interference; hence, $\hat{\mbf{x}}^{\NC}=[ \hat{x}^{\NC}_1\cdots\hat{x}^{\NC}_n\cdots\hat{x}^{\NC}_N ]$ is computed as
\begin{equation}\label{eq:SIC2}
\hat{x}^{\NC}_n=\left\lfloor \left(\tilde{y}_{n} - \sum_{l=n+1}^{N} r_{nl}\hat{x}^{\NC}_{l}\right)/r_{nn} \right\rceil_{\mathcal{M}},
\end{equation}
for $n = N,N-1,\cdots, 1$, where $\lfloor \alpha \rceil_{\mathcal{M}} \triangleq \argmin_{x \in \mathcal{M}} \abs{\alpha-x}$ is the slicing operator on the constellation $\mathcal{M}$. N/C serves as an upper bound on the performance of other detection schemes.

%
\subsection{Chase Detector (CD)}\label{sec:CD}
The CD \cite{waters2008chase} mitigates error propagation in SIC by populating a list $\mathcal{S}(\mbf{\tilde{y}},\mbf{R})$ of candidate symbol vectors for final decision. It first partitions $\mbf{\tilde{y}}$, $\mbf{R}$, and $\mbf{x}$ in~\eqref{eq:sysmodel2} as
\begin{equation}\label{eq:partitionCD}
    \mbf{\tilde{y}} =
        \begin{bmatrix}
            \mbf{\tilde{y}}_{1} \\
            \tilde{y}_{N}^{}
        \end{bmatrix}, \ \
    \mbf{R} =
        \begin{bmatrix}
            \mbf{A} & \mbf{b} \\
            \mbf{0} & c
        \end{bmatrix}, \ \
    \mbf{x} =
        \begin{bmatrix}
            \mbf{x}_{1} \\
            x_{N}^{}
        \end{bmatrix},
\end{equation}
where $\mbf{\tilde{y}}_{1} \in \mathcal{C}^{(N-1)\times1}$, $\tilde{y}_{N}^{} \in \mathcal{C}^{1\times1}$, $\mbf{A} \in \mathcal{C}^{(N-1)\times(N-1)}$, $\mbf{b}\in \mathcal{C}^{(N-1)\times1}$, $c \in \mathcal{R}^{1\times1}$, $\mbf{x}_{1} \in \mathcal{M}^{N-1}$, $\mbf{0}$ is a $1\times(N-1)$ vector of zero-valued entries, and $x_{N}^{} \in \mathcal{M}$.
Then, for each $x_N$ at the root layer, a candidate vector is calculated as in~\eqref{eq:SIC2} and added to $\mathcal{S}$. The maximum number of candidate vectors in $\mathcal{S}$ is $\abs{\mathcal{M}}$, and the final HO decision vector is chosen from $\mathcal{S}$ to be
\begin{equation}\label{eq:sol_CD}
  \hat{\mbf{x}}^{\CD} = \argmin_{\mbf{x} \in \mathcal{S}} \norm{\mbf{\tilde{y}}\!-\!\mbf{R}\mbf{x}}^{2}.
\end{equation}
Note that CD differs from LSD \cite{hochwald2003achieving} in several aspects. For example, LSD list admission depends on run-time channel conditions, which makes it nondeterministic and more complex. Also, in a SO setting, LSD does not guarantee computing all the required distance metrics.

%
\subsection{ Layered Orthogonal Lattice Detector (LORD)}\label{sec:LORD}
Instead of executing the CD routine once, LORD repeats chase detection with different layer orderings, each time with a different layer as root, by cyclically shifting the columns of $\mbf{H}$. The best output from these trials is the final solution. Each permuted $\mbf{H}$ at step $t$, $t\!=\!1,\!\cdots\!,N$, is QR-decomposed into $\mbf{Q}^{(t)}$ and $\mbf{R}^{(t)}$ according to~\eqref{eq:partitionCD}. Let $\hat{\mbf{x}}_{(t)}^{\CD}$ denote the output CD solution from step $t$. Then, the final solution $\hat{\mbf{x}}^{\LORD}$ is $\hat{\mbf{x}}_{(t_{min})}^{\CD}$, where
\begin{equation}\label{eq:LORDout}
    t_{min} = \argmin_{ t \in \{1,\cdots,N\} } \norm{\mbf{\tilde{y}} - \mbf{R}\hat{\mbf{x}}_{(t)}^{\CD}}^{2}.
\end{equation}
Since distances are preserved under different layer orderings with QRD, the accumulated candidate vectors across different partitions form an ``extended'' candidate list, despite the potential overlap of lists from each partition. Therefore, the added gain with LORD compared to CD is significant.

%
\section{Detection Schemes Based on Punctured Channel Matrix}\label{sec:proposed}

%
\subsection{Punctured QR Decomposition (WRD)}\label{sec:WRD}
WRD transforms $\mbf{H}$ into a punctured UTM $\Rp=[\rp_{ij}]\in\mathcal{C}^{N\times N}$ with $\rp_{ii}\in\mathcal{R}^{+}$, by puncturing entries between the diagonal and the last column through a matrix $\mbf{W}=[\mbf{w}_{1}\cdots \mbf{w}_{n} \cdots \mbf{w}_{N}^{}]\in\mathcal{C}^{M\times N}$, such that $\mbf{W}^{*}\mbf{H}=\Rp$. A brute force approach for computing $\mbf{W}$ \cite{ojard2008method} involves matrix inversions, which is complex and prone to roundoff error. However, an alternative approach that employs QRD followed by elementary matrix operations can be used to derive $\mbf{W}$ and $\Rp$~\cite{2014_mansour_eurasip_WLD}.

Let $\mbf{H}$ be QR-decomposed such that $\mbf{Q}^{*}\mbf{H}=\mbf{R}$. Obviously, $\mbf{q}_{N}^{*}\mbf{q}_{N}^{}=1$ and $\mbf{q}^{*}_{N}\mbf{h}_{n}=0$ for all $n=1,\ldots,N\!-\!1$, hence, $\mbf{w}_{N}^{}=\mbf{q}_{N}^{}$. Now assume the $\nth{n}$ entry $r_{mn}$ in row $m$ of $\mbf{R}$ is to be nulled, for $m\!=\!1,\cdots,N\!-\!2$ and $n\!=\!m\!+\!1,\cdots,N\!-\!1$. We have $\mbf{q}_{m}^{*}\mbf{h}_{n}=r_{mn}\in\mathcal{C}$ and $\mbf{q}_{n}^{*}\mbf{h}_{n}=r_{nn}\in\mathcal{R}^{+}$, from which it follows that $\left(\mbf{q}^{*}_{m}-\mbf{q}^{*}_{n}\frac{r_{mn}}{r_{nn}}\right)\mbf{h}_{n}=0$. Hence, with $\rho_{mn}^{}\!\triangleq\!\tfrac{r_{mn}}{r_{nn}} \in\mathcal{C}$, the equations
\begin{align}\label{eq:punc1}
    \mbf{q}_{m}^{} &= \mbf{q}_{m}^{}  -   \mbf{q}_{n}^{} \rho_{mn}^*, \\
    r_{mn}^{}      &= r_{mn}^{}       -   r_{nn}^{} \rho_{mn}^{},~~\text{and}     \\
    r_{mN}^{}      &= r_{mN}^{}       -   r_{nN}^{} \rho_{mn}^{},
\end{align}
when repeated for $n=N-1,N-2,\cdots,m+1$, would puncture the required $\nth{n}$ entry and update the $\nth{N}$ entry in row $m$ of $\mbf{R}$, as well as update the $\nth{m}$ column of $\mbf{Q}$ accordingly, while
\begin{align}\label{eq:punc2}
  r_{mm}^{} &= r_{mm}/\norm{\mbf{q}_{m}}, \\
  r_{mN}^{} &= r_{mN}/\norm{\mbf{q}_{m}},~~\text{and} \\
  \mbf{q}_{m} &= \mbf{q}_{m}/\norm{\mbf{q}_{m}},
\end{align}
would normalize $\mbf{q}_{m}$ in $\mbf{Q}$ and update the non-zero entries in row $m$ of $\mbf{R}$ accordingly. All these operations are to be carried for $m\!=\!N\!-\!2,N\!-\!3,\cdots,1$. The resultant $\mbf{Q}$ is $\mbf{W}$, and the resultant $\mbf{R}$ is $\Rp$. The transformed received symbol vector after applying $\mbf{W}^*$ can then be expressed as
\begin{equation}\label{eq:sysmodel3}
  \mbf{\bar{y}} = \mbf{W}^{*}\mbf{y} = \Rp\mbf{x} + \mbf{W}^{*}\mbf{n},
\end{equation}
such that
\begin{equation}\label{eq:partitionCDwrd}
\mbf{\bar{y}} =
    \left[\begin{IEEEeqnarraybox*}[][c]{c}
        \mbf{\bar{y}}_{1} \\
        \bar{y}_{N}^{}
    \end{IEEEeqnarraybox*}\right], \ \
\Rp =
    \begin{bmatrix}
        \mathring{\mbf{A}} & \mathring{\mbf{b}} \\
        \mbf{0} & \mathring{c}
        \end{bmatrix}, \ \  \mbf{x}=\begin{bmatrix} \mbf{x}_{1} \\ x_{N}^{}  \end{bmatrix},
\end{equation}
where in this case $\mathring{\mbf{A}} \in \mathcal{R}^{(N-1)\times(N-1)}$ is a diagonal matrix. For example, in the special case of $4\!\times\!4$ MIMO, $\Rp$ is obtained from $\mbf{R}$ by puncturing entries $r_{23}^{},r_{12}^{},r_{13}^{}$:
\begin{equation}\label{eq:SICmatrices}
  \mbf{R} =
    \begin{bsmallmatrix}
        r_{11}  & r_{12}     & r_{13}     & r_{14} \\
        0       & r_{22}     & r_{23}     & r_{24} \\
        0       & 0          & r_{33}     & r_{34} \\
        0       & 0          & 0          & r_{44}
    \end{bsmallmatrix},\quad
  \Rp =
    \begin{bsmallmatrix}
        \rp_{11} & 0 & 0 & \rp_{14} \\
        0 & \rp_{22} & 0 & \rp_{24} \\
        0 & 0 & \rp_{33} & \rp_{34} \\
        0 & 0 & 0 & \rp_{44}
    \end{bsmallmatrix}.
\end{equation}

\begin{figure*}[t]
  \centering
  \subfloat[]{\label{distfig:a} \includegraphics[width=0.48\linewidth]{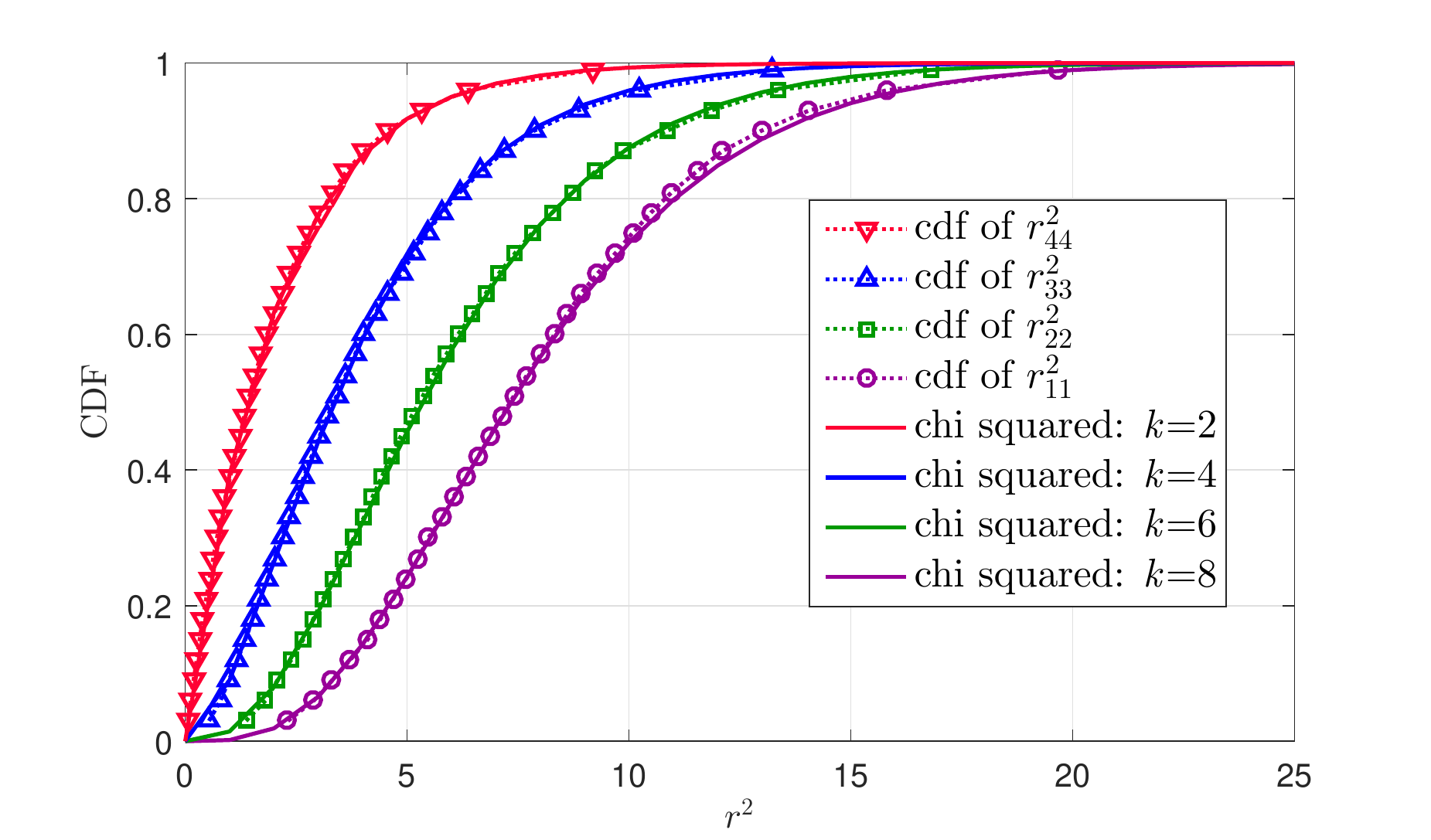}}
  \hfill
  \subfloat[]{\label{distfig:b} \includegraphics[width=0.48\linewidth]{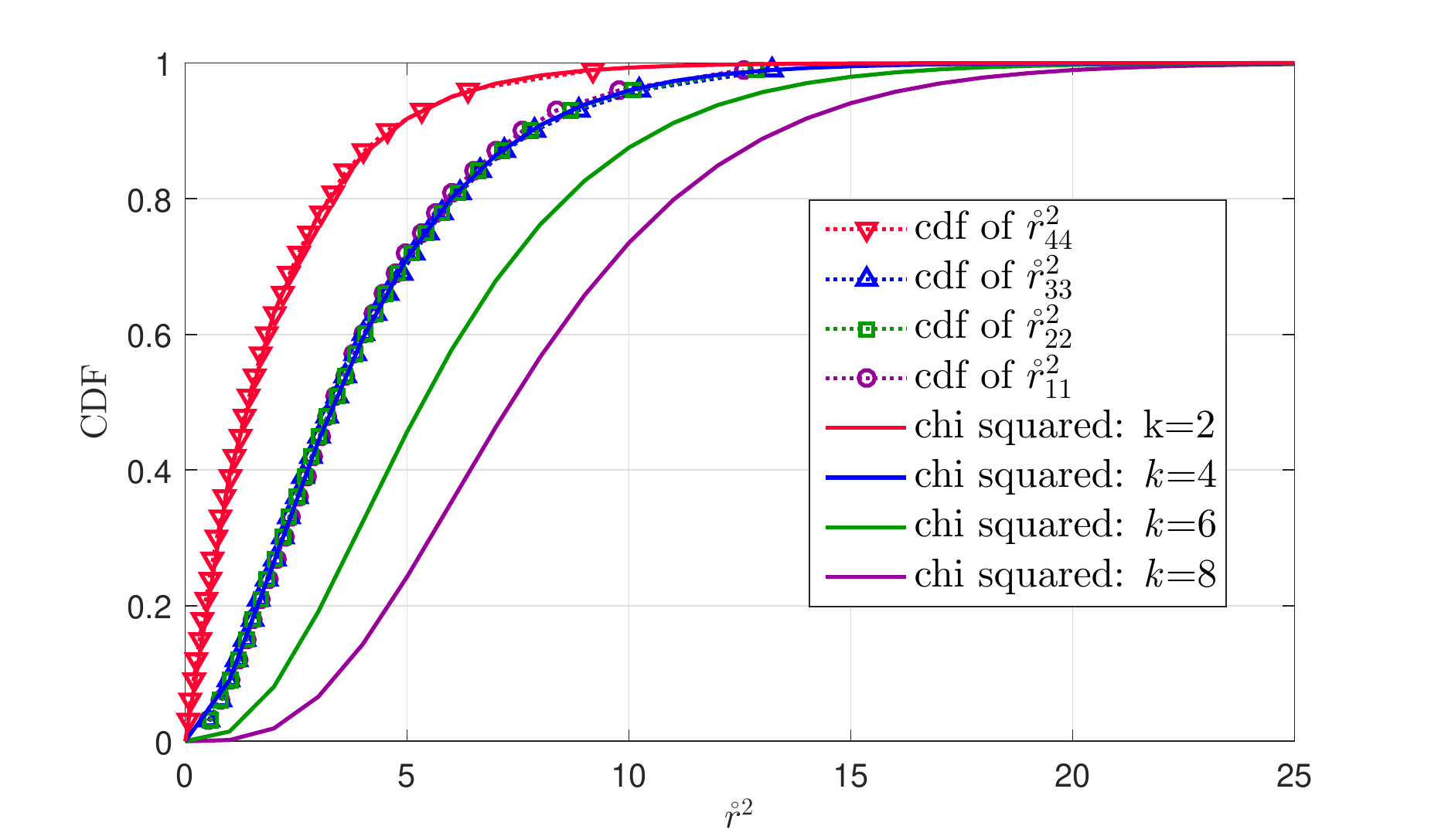}}
  \caption{Empirical cumulative distribution functions (CDFs) of the diagonal elements of (a) $\mbf{R}$, and (b) $\Rp$ shown in dotted lines compared to theoretical chi-squared CDFs in solid lines.}
  \label{f:distribution}
\end{figure*}

Note that the column at the root layer in $\mbf{W}$ (layer $N$ here), remains orthogonal to all other columns. Hence, taking the expectation of $\mbf{W}^*\mbf{nn}^*\mbf{W}$ over $\mbf{n}$, we have:
\begin{equation}\label{eq:matrix}
    \mathsf{E}_{\mbf{n}}[\mbf{W}^*\mbf{nn}^*\mbf{W}] =
        \begin{bsmallmatrix}
            \sigma^2    & e_{12}    & e_{13}    & 0 \\
            e_{12}^*    & \sigma^2  & e_{23}    & 0 \\
            e_{13}^*    & e_{23}^*  & \sigma^2  & 0 \\
            0           & 0         & 0         & \sigma^2
        \end{bsmallmatrix}.
\end{equation}
Therefore, although the resultant noise after puncturing is colored, WRD preserves the noise variance at the layer of interest. However, the statistical properties of the elements of $\Rp$ get distorted under puncturing. The non-zero elements of $\mbf{R}$ (given i.i.d. Rayleigh fading) are known to be independent random variables with the following distributions~\cite{edelman1988eigenvalues,choi2006nulling}:
\begin{itemize}
  \item The off-diagonal elements are circular symmetric complex Gaussian with unit variance.
  \item The square of the $\nth{n}$ diagonal element is chi-squared distributed with $k=2(N-n+1)$ degrees of freedom, and its probability density function is given by
      \begin{equation}
      f(g=r_{nn}^2) = \frac{1}{(N-n)!} g^{N-n}e^{-g}, \ \ g \geq 0,
      \end{equation}
\end{itemize}
where chi-squared comes from the sum of squares of Rayleigh distributed random variables. While the distributions of non-zero off-diagonal elements remain intact, the distributions of diagonal elements at upper layers $n\!=\!1,\cdots,N\!-\!3$, lose degrees of freedom from $2(N\!-\!n\!+\!1)$ down to $4$, as depicted in Fig.~\ref{f:distribution} for a $4\times4$ channel matrix. This is caused by the fact that each puncturing operation at layer $n$ renders the $\nth{n}$ column of $\mbf{W}$ dependent on one of the remaining columns, thus eliminating two degrees of freedom from the corresponding distribution of $\rp_{nn}^2$.

%
Similar to the ML detector, an ``exhaustive'' PML detector searches $\mathcal{X}$ to find
\begin{equation}\label{eq:PML_dist}
   \hat{\mbf{x}}^{\PML} = \min_{\mbf{x} \in \mathcal{X}}\norm{\mbf{W}^*(\mbf{y} - \mbf{H}\mbf{x})}^{2}.
\end{equation}
Pre-multiplying by $\mbf{W}$, unlike $\mbf{Q}$, modifies Euclidean distances, hence we have
\begin{equation}\label{eq:mod_dist}
d(\mbf{x}) = \norm{\mbf{y} - \mbf{H}\mbf{x}}^{2} = \norm{\mbf{Q}^*(\mbf{y} - \mbf{H}\mbf{x})}^{2} \neq \norm{\mbf{W}^*(\mbf{y} - \mbf{H}\mbf{x})}^{2} = \norm{\mbf{\bar{y}}\!-\!\Rp\mbf{x}}^{2} = \bar{d}(\mbf{x}).
\end{equation}
Note that this minimum distance detector is not optimal due to the presence of colored noise.

%
\subsection{Punctured N/C Detector (PN/C)}\label{sec:propSIC}
With PN/C, we null by pre-multiplying by $\mbf{W}^{*}$ instead of $\mbf{Q}^{*}$, and perform SIC as
\begin{equation}\label{eq:SICpunctured}
    \hat{x}^{\PNC}_{n} = \left\lfloor \left(\bar{y}_{n} - \rp_{nN}^{}\hat{x}_{N}^{\PNC}\right)/\rp_{nn} \right\rceil_{\mathcal{M}},
\end{equation}
for $n = N-1,\cdots, 1$, where $\hat{\mbf{x}}^{\PNC}=[ \hat{x}^{\PNC}_1\cdots\hat{x}^{\PNC}_n\cdots\hat{x}^{\PNC}_N ]$, and $\hat{x}^{\PNC}_{N} = \left\lfloor \bar{y}_{N}/\rp_{NN}\right\rceil_{\mathcal{M}}$. Note that slicing on layers $n = N-1,\cdots, 1$ can be done in parallel since $\mathring{\mbf{A}}$ is diagonal.

%
\subsection{Punctured Chase Detector (PCD)}\label{sec:propCD}
The PCD builds on the partition in equation \eqref{eq:partitionCDwrd}, and performs the operations of a CD (Sec. \ref{sec:CD}). A modified list of candidate symbol vectors $\mathcal{P}(\mbf{\bar{y}},\Rp)$ is thus created. The distance of a vector $\mbf{x}=[\mbf{x}_1,x_N]^T$ is given by
\begin{equation}\label{eq:xwr}
    \bar{d}(\mbf{x})    \!=\!
        \norm{\mbf{\bar{y}}\!-\!\Rp\mbf{x}}^{2} \!=\! \abs{\bar{y}_{N}^{}\!-\!\mathring{c} x_{N}^{}}^{2}\!+\!\norm{\mbf{\bar{y}}_{1}\!-\!\mathring{\mbf{A}}\mbf{{x}}_{1}\!-\!\mathring{\mbf{b}}x_{N}^{}}^{2}.
\end{equation}
For a given $x_N^{}\in\mathcal{M}$, the distance in~\eqref{eq:xwr} is minimized as
\begin{align}\label{eq:xwr2}
    \min_{\mbf{x}_1\in \mathcal{M}^{N-1}} \bar{d}(\mbf{x})
        &= \abs{\bar{y}_{N}^{}\!-\!\mathring{c} x_{N}^{}}^{2}\!+\!
              \min_{\mbf{x}_1\in \mathcal{M}^{N-1}} \norm{\mbf{\bar{y}}_{1}\!-\!\mathring{\mbf{A}}\mbf{{x}}_{1}\!-\!\mathring{\mbf{b}}x_{N}^{}}^{2}\\
        &= \abs{\bar{y}_{N}^{}\!-\!\mathring{c} x_{N}^{}}^{2} +
              \norm{\mbf{\bar{y}}_{1}\!-\!\mathring{\mbf{A}}\mbf{\hat{x}}_{1}(x_N)\!-\!\mathring{\mbf{b}}x_{N}^{}}^{2}\\
        &\triangleq \bar{d}^*\left(\mbf{x}(x_N)\right),
\end{align}
where $\mbf{\hat{x}}_{1}(x_N)=\lfloor(\mbf{\bar{y}}_{1}\!-\!\mathring{\mbf{b}} x_{N}^{})/\mathring{\mbf{A}}\rceil_{\mathcal{M}^{N-1}}$, which is a vectorized slicing operation, and $\mbf{x}(x_N)=[\mbf{\hat{x}}_1(x_N),x_N]^T$. The symbol vector $\mbf{x}(x_N)$ is then added to $\mathcal{P}$, together with its distance $\bar{d}^*\left(\mbf{x}(x_N)\right)$. The final HO symbol vector $\hat{\mbf{x}}^{\PCD}$ is found from $\mathcal{P}$ as the one with smallest distance.

While the PCD computes distances only to $\abs{\mathcal{P}(\mbf{\tilde{y}},\Rp)}=\abs{\mathcal{M}}$ candidate symbol vectors, for a given layer ordering and channel partition, it is clear from~\eqref{eq:xwr2} that it achieves the exact performance as that of the PML detector. In other words, there is no vector in the lattice $\mathcal{X}$, outside the set $\mathcal{P}(\mbf{\tilde{y}},\mbf{R})$, that can have a smaller distance metric than that of the PCD solution. The proof goes as follows:
\begin{align}\label{eq:xwr3}
    \min_{\mbf{x}\in \mathcal{X}} \bar{d}(\mbf{x})
        &=  \min_{x_N\in\mathcal{M},\mbf{x}_1\in \mathcal{M}^{N-1}}
            \left\{
            \abs{\bar{y}_{N}^{}\!-\!\mathring{c} x_{N}^{}}^{2}\!+\!
              \norm{\mbf{\bar{y}}_{1}\!-\!\mathring{\mbf{A}}\mbf{{x}}_{1}\!-\!\mathring{\mbf{b}}x_{N}^{}}^{2}
            \right\}\\
        &= \min_{x_N\in\mathcal{M}}
            \left\{
                \abs{\bar{y}_{N}^{}\!-\!\mathring{c} x_{N}^{}}^{2}\!+\!
                \min_{\mbf{x}_1\in \mathcal{M}^{N-1}} \norm{\mbf{\bar{y}}_{1}\!-\!\mathring{\mbf{A}}\mbf{{x}}_{1}\!-\!\mathring{\mbf{b}}x_{N}^{}}^{2}
            \right\}\\
        &= \min_{x_N\in\mathcal{M}}
            \left\{
                \abs{\bar{y}_{N}^{}\!-\!\mathring{c} x_{N}^{}}^{2} +
                \norm{\mbf{\bar{y}}_{1}\!-\!\mathring{\mbf{A}}\mbf{\hat{x}}_{1}(x_N)\!-\!\mathring{\mbf{b}}x_{N}^{}}^{2}
            \right\}   \\
        &= \min_{\mbf{x}(x_N)\in\mathcal{P}} \bar{d}^*\left(\mbf{x}(x_N)\right).
\end{align}

%
\subsection{Vector-Based Sub-Space Detector (VSSD)}\label{sec:propSSvb}
\begin{figure}[t]
\centering
\includegraphics[width=3.5in]{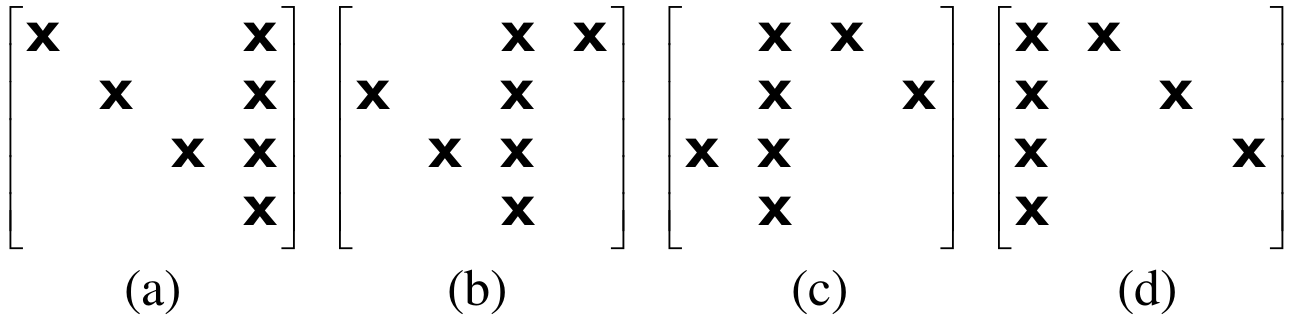}
\caption{Punctured $4\times 4$ channel decomposition structures under different permutations: (a) $t=1$; (b) $t=2$; (c) $t=3$; (d) $t=4$.}
\label{f:matrices}
\end{figure}

The VSSD is an extension to PCD, the same way LORD is an extension to CD. The columns of $\mbf{H}$ are cyclically shifted, and punctured UTMs are generated as shown in Fig. \ref{f:matrices}. Each permuted $\mbf{H}$ at step $t$, $t = 1,\cdots,N$, is WR-decomposed into $\mbf{W}^{(t)}$ and $\Rp^{(t)}$ according to~\eqref{eq:partitionCDwrd}. Let $\hat{\mbf{x}}_{(t)}^{\PCD}$ denote the PCD solution from step $t$. The final solution $\hat{\mbf{x}}^{\VSSD}$ is $\hat{\mbf{x}}_{(t_{min})}^{\PCD}$, where $t_{min}$ is defined as:
\begin{equation}\label{eq:SSDout}
t_{\min} = \argmin_{ t \in \{1,\cdots,N\} } \norm{\mbf{y} - \mbf{H}\hat{\mbf{x}}_{(t)}^{\PCD}}^{2}.
\end{equation}
Note that we revert back to the original space of $\mbf{H}$ to compute the true Euclidean distance metrics in~\eqref{eq:SSDout}. The gain achieved by VSSD compared to PCD is limited, since each $\Rp^{(t)}$ generates an independent space, and hence we end up taking the best output from $N$ independent trials. The VSSD is in effect the HO version of the reference SO SSD \cite{2014_mansour_SPL_WLD}, and we refer to it by simply SSD in the remainder of this paper.

%
\subsection{Symbol-Based Sub-Space Detector (SSSD)}\label{sec:propSSsb}
As a variation of SSD, the SSSD selects at each step $t$, only the root symbol of the output vector as a component of the final output vector. Thus, the output vector $\hat{\mbf{x}}^{\SSSD}=[ \hat{x}^{\SSSD}_1\cdots\hat{x}^{\SSSD}_n\cdots\hat{x}^{\SSSD}_N ]$ gets assembled one symbol at a time over $N$ executions of PCD, where
\begin{equation}\label{eq:SSSDout}
\hat{x}^{\SSSD}_n = \hat{x}_{N\!-\!n\!+\!1(t=n)}^{\PCD}.
\end{equation}
For example, in a $4\!\times\!4$ MIMO system, we have $\hat{x}^{\SSSD}_1 = \hat{x}_{4(t=1)}^{\PCD}$, where $\hat{\mbf{x}}_{(t=1)}^{\PCD}$ is the HO solution of a PCD following the partition in Fig. \ref{f:matrices}(a). Similarly $\hat{x}^{\SSSD}_2 = \hat{x}_{3(t=2)}^{\PCD}$, $\hat{x}^{\SSSD}_3 = \hat{x}_{2(t=3)}^{\PCD}$, and $\hat{x}^{\SSSD}_4 = \hat{x}_{2(t=4)}^{\PCD}$, are obtained following the partitions (b), (c), and (d), respectively. Note that we can define symbol-based LORD (SLORD) in a similar manner:
\begin{equation}\label{eq:SLORD}
\hat{\mbf{x}}^{\SLORD}=[\hat{x}^{\SLORD}_1 \cdots \hat{x}^{\SLORD}_n \cdots\hat{x}^{\SLORD}_N];~~ \hat{x}^{\SLORD}_n = \hat{x}_{N\!-\!n\!+\!1(t=n)}^{\CD}.
\end{equation}

%
\section{Analysis of Achievable Diversity Gain}\label{sec:diversity}
It is known that ML detection achieves full receive diversity $M$, and it can be shown that the N/C and PN/C detectors, being special cases of ZF with decision feedback, can only achieve a receive diversity gain of $1$. Moreover, it can be argued that both SSD (VSSD) and LORD also achieve full diversity, since they exploit the full channel matrix $\mbf{H}$ to compute distance metrics. In what follows, we study the achievable diversity gains of PML (PCD), SSSD, and SLORD.

%
\subsection{Punctured ML Detector / Punctured Chase Detector (PML/PCD)}
\label{sec:diversityCD}
To capture the diversity order of PML, we derive the pairwise error probability (PEP). Suppose that $\mbf{x}^{(1)}$ is transmitted, while $\mbf{x}^{(2)}$ is erroneously detected, the PEP can be expressed as \\
\begin{align}\label{eq:app2_1}
\Prb(\mbf{x}^{(1)} \rightarrow \mbf{x}^{(2)}) & = \Prb\left(\norm{\mbf{W}^*(\mbf{y} - \mbf{H}\mbf{x}^{(2)})}^{2} \leq \norm{\mbf{W}^*(\mbf{y} - \mbf{H}\mbf{x}^{(1)})}^{2} \right)\\
 & = \Prb\left(\norm{\mbf{W}^*\mbf{H}(\mbf{x}^{(1)} - \mbf{x}^{(2)}) + \mbf{W}^*\mbf{n}}^{2} \leq \norm{\mbf{W}^*\mbf{n}}^{2} \right) \\
 & = \Prb\left(\Re(\mbf{n}^*\mbf{W}\Rp\mbf{d}) \geq \tfrac{1}{2}\norm{\Rp\mbf{d}}^{2} \right),
\end{align}
where $\Prb(\alpha)$ is the probability that event $\alpha$ occurs, and $\mbf{d} \triangleq \mbf{x}^{(1)} - \mbf{x}^{(2)}$. Since $\mbf{n}$ consists of circular symmetric complex Gaussian random variables, then so is $\mbf{n}^*\mbf{W}\Rp\mbf{d}$. It is easy to show that
\begin{align}\label{eq:app2_2}
    \mathsf{E}\left[\mbf{n}^*\mbf{W}\Rp\mbf{d}\right] & = 0 \\
    \mathsf{E}\left[(\mbf{n}^*\mbf{W}\Rp\mbf{d})(\mbf{n}^*\mbf{W}\Rp\mbf{d})^*\right] & = \mathsf{E}\left[\Tr(\Rp^*\mbf{W}^*\mbf{n}\mbf{n}^*\mbf{W}\Rp\mbf{d}\mbf{d}^*)\right]= \sigma^{2} \Tr\left( (\mbf{W}\Rp)^*(\mbf{W}\Rp)\mbf{d}\mbf{d}^* \right),
\end{align}
where $\Tr(\cdot)$ is introduced since $(\mbf{n}^*\mbf{W}\Rp\mbf{d})(\mbf{n}^*\mbf{W}\Rp\mbf{d})^*$ is a scalar. Hence, we have
\begin{align*}
    \Re(\mbf{n}^*\mbf{W}\Rp\mbf{d}) & \sim \mathcal{N} \left( 0, \frac{\sigma^{2}}{2} \Tr \left( (\mbf{W}\Rp)^*(\mbf{W}\Rp)\mbf{d}\mbf{d}^* \right) \right) = \mathcal{N} \left( 0, \frac{\sigma^{2}}{2} \norm{\mbf{W}\Rp\mbf{d}}^2 \right),
\end{align*}
and therefore,
\begin{align}\label{eq:app2_5}
\Prb(\mbf{x}^{(1)} \rightarrow \mbf{x}^{(2)}) & = Q\left( \frac{\norm{\Rp\mbf{d}}^2}{\sqrt{2\sigma^{2}}\norm{\mbf{W}\Rp\mbf{d}} }  \right) \leq Q\left( \sqrt{\frac{\norm{\Rp\mbf{d}}^2}{2\sigma^{2}\norm{\mbf{W}}^2_{\F}}}  \right),
\end{align}
where the inequality holds since $\norm{\mbf{W}\Rp\mbf{d}}\!\leq\!\norm{\mbf{W}}_{\F}\norm{\Rp\mbf{d}}$ (section 5.2 in \cite{meyer2000matrix}). Moreover, using union bound, we have
\begin{align}\label{eq:app2_6}
\Prb(\mbf{x}^{(1)} \rightarrow \mbf{x}^{(2)}) & \leq Q\left( \sqrt{\frac{\norm{\Rp\mbf{d}_{\min}}^2}{2\sigma^{2}\norm{\mbf{W}}^2_{\F}}}  \right) \leq \sum_{\mbf{d}\in \Omega,\mbf{d}\neq0} Q\left( \sqrt{\frac{\norm{\Rp\mbf{d}}^2}{2\sigma^{2}\norm{\mbf{W}}^2_{\F}}}  \right),
\end{align}
where $\Omega\triangleq\{\mbf{d}\!=\!\mbf{x}\!-\!\acute{\mbf{x}} \ |\  \mbf{x},\acute{\mbf{x}}\!\in\!\mathcal{X} \}$, and $\mbf{d}_{\min}\!=\!\argmin_{\mbf{d}\in \Omega,\mbf{d}\neq0} \norm{\Rp\mbf{d}}^2$. Finally, using the Chernoff bound, the average PEP is upper bounded as
\begin{equation}\label{eq:app2_7}
\mathsf{E}\left[\Prb(\mbf{x}^{(1)} \rightarrow \mbf{x}^{(2)})\right] \leq \sum_{\mbf{d}\in \Omega,\mbf{d}\neq0} \mathsf{E}\left[ \exp \left(-\frac{\norm{\Rp\mbf{d}}^2}{4\sigma^{2}\norm{\mbf{W}}^2_{\F}}\right)  \right] = \sum_{\mbf{d}\in \Omega,\mbf{d}\neq0} \mathsf{E}\left[ \exp \left(-\frac{\norm{\Rp\mbf{d}}^2}{4N\sigma^{2}}\right)  \right],
\end{equation}
where $\norm{\mbf{W}}^2_{\F} = N$ since the columns of $\mbf{W}$ were normalized in \eqref{eq:punc2}.

For regular ML detection \cite{Biglieri_2002,bai2014low,larsson2008space}, we have
\begin{equation}\label{eq:app2_8}
\mathsf{E}\left[\Prb(\mbf{x}^{(1)} \rightarrow \mbf{x}^{(2)})\right] \leq \sum_{\mbf{d}\in \Omega,\mbf{d}\neq0} \mathsf{E}\left[ \exp \left(-\frac{\norm{\mbf{H}\mbf{d}}^2}{4\sigma^{2}}\right)  \right] \leq \sum_{\mbf{d}\in \Omega,\mbf{d}\neq0} \det \left( \mbf{I}_N + \frac{\mbf{d}\mbf{d}^*}{4\sigma^{2}} \right)^{-M},
\end{equation}
where the expected value over the elements of $\mbf{H}$ results in full receive diversity $M$, because each column of $\mbf{H}$ contains $M$ independent Rayleigh distributed random variables, whose square is exponentially distributed. However, with $\Rp$ instead of $\mbf{H}$ in PML detection, the first $N\!-\!1$ columns have single diagonal elements, whose squares are chi-squared distributed with $4$ degrees of freedom, which corresponds to two exponentially distributed complex random variables, and hence a receive diversity order equal to $2$. Only column $N$ of $\Rp$ provides a diversity equal to $M$. Therefore, by analogy with \eqref{eq:app2_8}, the average PEP for the PML detector is
\begin{equation}\label{eq:app2_9}
\mathsf{E}\left[\Prb(\mbf{x}^{(1)} \rightarrow \mbf{x}^{(2)})\right] \leq \sum_{\mbf{d}\in \Omega,\mbf{d}\neq0} \det \left( \mbf{I}_N + \frac{\mbf{d}\mbf{d}^*}{4N\sigma^{2}} \right)^{-2},
\end{equation}
and hence PML detection can not achieve a receive diversity gain of order greater than 2. However, noting that PML and PCD are identical (Sec.~\ref{sec:propCD}), and knowing that the regular CD achieves a receive diversity order of 2 (more on that in Sec.~\ref{sec:berCD}), we conclude that channel puncturing does not reduce the diversity gain of the CD.

%
\subsection{Symbol-Based Sub-Space Detector (SSSD)}
\label{sec:diversitySSSD}

To capture the diversity order of SSSD, we derive a modified PEP. Without loss of generality, we assume that layer $N$ is the root layer of interest. Hence, an error occurs when $x_N^{(1)}$ is transmitted and $x_N^{(2)}$ is erroneously detected, with probability
\begin{align}\label{eq:app3_1}
\Prb(x_N^{(1)} \rightarrow x_N^{(2)}) & = \Prb\left(\norm{\mbf{W}^*\left(\mbf{y} - \mbf{H}_1\mbf{x}_1^{(2)} - \mbf{h}_N x_N^{(2)}\right)}^{2} \leq \norm{\mbf{W}^*\left(\mbf{y} - \mbf{H}_1\mbf{x}_1^{(1)} - \mbf{h}_N x_N^{(1)}\right)}^{2} \right)\\
 & = \Prb\left(\norm{\mbf{W}^*\mbf{H}_1\left(\mbf{x}_1^{(1)} - \mbf{x}_1^{(2)}\right) + \mbf{W}^*\mbf{h}_N\left(x_N^{(1)} - x_N^{(2)}\right) + \mbf{W}^*\mbf{n}}^{2} \leq \norm{\mbf{W}^*\mbf{n}}^{2} \right) \\
 & = \Prb\left(\norm{\Rp_{1}\left(\mbf{x}_1^{(1)} - \mbf{x}_1^{(2)}\right) + \rrp_{N}\left(x_N^{(1)} - x_N^{(2)}\right) + \mbf{W}^*\mbf{n}}^{2} \leq \norm{\mbf{W}^*\mbf{n}}^{2} \right),
\end{align}
where $\mbf{y} = \mbf{H}_1\mbf{x}_1^{(1)} + \mbf{h}_N x_N^{(1)} + \mbf{n}$, $\mbf{x}_1^{(2)} = \mbf{\hat{x}}_{1}(x_N^{(2)})$ is computed as in Sec.~\ref{sec:propCD}, $\mbf{h}_N$ and $\rrp_{N}$ are the \nth{\emph{N}} column of $\mbf{H}$ and $\Rp$, and $\mbf{H}_1\in\mathcal{C}^{M\times (N-1)}$ and $\Rp_{1}\in\mathcal{C}^{M\times (N-1)}$ are the first $N-1$ columns of $\mbf{H}$ and $\Rp$, respectively. Let $\mathring{\mbf{\Delta}} = \Rp_{1} \left(\mbf{x}_1^{(1)} - \mbf{x}_1^{(2)}\right)$, and let $d = x_N^{(1)} - x_N^{(2)}$; we have
\begin{align}\label{eq:app3_2}
\Prb\left(x_N^{(1)} \rightarrow x_N^{(2)}\right) & = \Prb\left(\norm{\rrp_{N} d + \left(\mbf{W}^*\mbf{n} + \mathring{\mbf{\Delta}}\right)}^{2} \leq \norm{\mbf{W}^*\mbf{n}}^{2} \right)\notag \\
& = \Prb\left(\norm{\rrp_{N} d}^{2} \leq -2\Re\left(\left(\mbf{W}^*\mbf{n} + \mathring{\mbf{\Delta}}\right)^*\rrp_{N} d\right) - \norm{\mbf{W}^*\mbf{n} + \mathring{\mbf{\Delta}}}^{2} + \norm{\mbf{W}^*\mbf{n}}^{2} \right) \notag\\
 & \leq \Prb\left( -2\Re\left(\left(\mbf{W}^*\mbf{n} + \mathring{\mbf{\Delta}}\right)^*\rrp_{N} d\right) \geq  \norm{\rrp_{N} d}^{2} - \norm{\mbf{W}^*\mbf{n}}^{2} \right).
\end{align}
Since $\mbf{n}$ and the columns of $\mbf{H}$ are circular symmetric complex Gaussian, then so is $\mbf{n}^*\mbf{W}\rrp_{N} d + \mathring{\mbf{\Delta}}^*\rrp_{N} d$. Thus, it can be shown that
\begin{align}\label{eq:app3_3}
    \mathsf{E}\left[(\mbf{W}^*\mbf{n} + \mathring{\mbf{\Delta}})^*\rrp_{N} d\right] & = 0 \\
    \mathsf{E}\left[\left(\left(\mbf{W}^*\mbf{n} + \mathring{\mbf{\Delta}}\right)^*\rrp_{N} d\right)\left(\left(\mbf{W}^*\mbf{n} + \mathring{\mbf{\Delta}}\right)^*\rrp_{N} d\right)^*\right] & = \sigma^{2}\norm{\mbf{W}\rrp_{N} d}^2 + \norm{\mathring{\mbf{\Delta}}^*\rrp_{N} d}^2 \\
    \Re\left(\left(\mbf{W}^*\mbf{n} + \mathring{\mbf{\Delta}}\right)^*\rrp_{N} d\right) & \sim \mathcal{N} \left( 0, \frac{\sigma^{2}}{2}\norm{\mbf{W}\rrp_{N}d}^2 + \frac{1}{2}\norm{\mathring{\mbf{\Delta}}^*\rrp_{N}d}^2   \right).
\end{align}
Hence, continuing from~\eqref{eq:app3_2}, we have
\begin{align}\label{eq:app3_6}
\Prb\left(x_N^{(1)} \rightarrow x_N^{(2)}\right) &\leq Q\left( \frac{\norm{\rrp_{N} d}^{2} - \norm{\mbf{W}^*\mbf{n}}^{2}} {\sqrt{2\sigma^{2}\norm{\mbf{W}\rrp_{N}d}^2 + 2\norm{\mathring{\mbf{\Delta}}^*\rrp_{N}d}^2} }  \right) \\
 & = Q\left( \sqrt{\frac{ \norm{\rrp_{N} d}^{4} - 2\norm{\rrp_{N} d}^{2}\norm{\mbf{W}^*\mbf{n}}^{2} +\norm{\mbf{W}^*\mbf{n}}^{4} } {2\sigma^{2}\norm{\mbf{W}\rrp_{N}d}^2 + 2\norm{\mathring{\mbf{\Delta}}^*\rrp_{N}d}^2 }}  \right) \\
 & \leq Q\left( \sqrt{\frac{ \norm{\rrp_{N} d}^{4} - 2\norm{\rrp_{N} d}^{2}\norm{\mbf{W}^*\mbf{n}}^{2} } {\norm{\rrp_{N}d}^2 \left( 2\sigma^{2}\norm{\mbf{W}}^2_{\F} + 2\norm{\mathring{\mbf{\Delta}}}^2 \right)}}  \right) \\
 & = Q\left( \sqrt{\frac{ \norm{\rrp_{N} d}^{2} - 2\norm{\mbf{W}^*\mbf{n}}^{2} } {2\sigma^{2}\norm{\mbf{W}}^2_{\F} + 2\norm{\mathring{\mbf{\Delta}}}^2}}  \right) = Q\left( \sqrt{\frac{ d^2\norm{\rrp_{N}}^{2} - 2N\sigma^{2} } {2N\sigma^{2} + 2\norm{\mathring{\mbf{\Delta}}}^2}}  \right).
\end{align}
Then, using union and Chernoff bounds, with $\Phi\triangleq\{d\!=\!x\!-\!\acute{x} \ |\  x,\acute{x}\!\in\!\mathcal{M} \}$ ($\abs{d}^2\!=\!dd^*$), the average PEP can be upper bounded as \\
\begin{align}\label{eq:app3_7}
\mathsf{E}\left[\Prb \left(x_N^{(1)} \rightarrow x_N^{(2)}\right)\right] & \leq \sum_{d\in \Phi,d\neq0} \mathsf{E}\left[ \exp \left(-\frac{ \abs{d}^2\norm{\rrp_{N}}^{2} - 2N\sigma^{2} } {4N\sigma^{2} + 4\norm{\mathring{\mbf{\Delta}}}^2}\right)  \right] \\
& = \sum_{d\in \Phi,d\neq0} \mathsf{E}\left[ \exp \left(-\frac{ \abs{d}^2\norm{\rrp_{N}}^{2} } {4N\sigma^{2} + 4\norm{\mathring{\mbf{\Delta}}}^2}\right) \exp \left(\frac{ 2N\sigma^{2} } {4N\sigma^{2} + 4\norm{\mathring{\mbf{\Delta}}}^2}\right)  \right] \\
& \approx \sum_{d\in \Phi,d\neq0} \mathsf{E}\left[ \exp \left(-\frac{ \abs{d}^2\norm{\rrp_{N}}^{2} } {4N\sigma^{2} + 4\norm{\mathring{\mbf{\Delta}}}^2}\right) \right],
\end{align}
where the last approximation holds since the second exponential term is less that $\exp(1)$, with equality at high $\mathsf{SNR}$ ($\sigma^{2}\!=\!0)$. Finally, taking the expectation over all squared elements of $\rrp_{N}$, which are exponentially distributed, we obtain
\begin{equation}\label{eq:app3_8}
\mathsf{E}\left[\Prb \left(x_N^{(1)} \rightarrow x_N^{(2)}\right)\right] \leq \sum_{d\in \Phi,d\neq0} \prod_{l=1}^{M} \left( \frac{\abs{d}^2}{4N\sigma^{2} + 4\norm{\mathring{\mbf{\Delta}}}^2} +1 \right)^{-1} \leq \sum_{d\in \Phi,d\neq0} \left( \frac{\abs{d}^2}{4N\sigma^{2} + 4\norm{\mathring{\mbf{\Delta}}}^2} \right)^{-M}.
\end{equation}

The denominator $4N\sigma^{2} \!+\! 4\norm{\mathring{\mbf{\Delta}}}^2$ represents noise plus interference, hence, SSSD appears to achieve a full receive diversity gain at the layer of interest when BERs are plotted in terms of signal-to-interference-plus-noise ratio ($\mathsf{SINR}$). In the case of SLORD, following a similar derivation, the average PEP can be expressed as
\begin{equation}\label{eq:app3_9}
\mathsf{E}\left[\Prb \left(x_N^{(1)} \rightarrow x_N^{(2)}\right)\right] \leq \sum_{d\in \Phi,d\neq0}  \left( \frac{\abs{d}^2}{4N\sigma^{2} + 4\norm{\mbf{\Delta}}^2} \right)^{-M},
\end{equation}
where $\mbf{\Delta} = \mbf{R}_{1} \left(\mbf{x}_1^{(1)} - \mbf{x}_1^{(2)}\right)$, and $\mbf{R}_1\in\mathcal{C}^{M\times (N-1)}$ consists of the first $N-1$ columns of $\mbf{R}$. Note that $\mathring{\mbf{\Delta}}$ can be expressed as
\begin{equation}\label{eq:app3_10}
\mathring{\mbf{\Delta}} = \left[ \rp_{11}\left(x_1^{(1)}-x_1^{(2)}\right),\ \ldots \ , \rp_{(N\!-\!1)(N\!-\!1)}\left(x_{N\!-\!1}^{(1)}-x_{N\!-\!1}^{(2)}\right), 0\right]^T,
\end{equation}
and consequently, $\norm{\mathring{\mbf{\Delta}}}^2$ is upper bounded by
\begin{equation}\label{eq:app3_11}
\norm{\mathring{\mbf{\Delta}}}^2_{\max} = \rp_{11}^2\beta^2 + \ \cdots \ + \rp_{(N\!-\!1)(N\!-\!1)}^2\beta^2,
\end{equation}
when a one-bit slicing error (assuming Gray mapping) occurs on all upper layers, with $\beta = 2/\log_2(L)$ for an $L$-QAM constellation. Since the expected values of the chi-squared-distributed square of diagonal elements in $\mbf{R}$ are greater than those in $\Rp$ (expected value equals degrees of freedom), we have $\norm{\mbf{\Delta}}^2 > \norm{\mathring{\mbf{\Delta}}}^2$, and hence SSSD outperforms SLORD. Furthermore, with higher order constellations, $\norm{\mathring{\mbf{\Delta}}}^2_{\max}$ is significantly reduced, boosting the performance of SSSD.

%
\section{Characterization and Analysis of BER}
\label{sec:probBER}

%
\subsection{Punctured N/C Detector (PN/C)}
\label{sec:berSIC}
Let $P_n^{}(r_{nn}^{})$ and $\acute{P}_n^{}(\rp_{nn}^{})$ be the probabilities of bit error conditioned on $r_{nn}$ and $\rp_{nn}$, when detecting $x_n$ ($1\!\leq\!n\!\leq\!N$), for N/C and PN/C, respectively. Consider the PN/C detector, and assume as in \cite{choi2006nulling} normalized binary phase shift keying (BPSK), where $\mathcal{M}=\{-1,1\}$, we have
\begin{equation}\label{eq:qfun1}
    \acute{P}_N^{}(\rp_{NN}^{}) = Q \left( \sqrt{ \frac{ 2\rp_{NN}^2 }{ \sigma^{2} } } \right)
\end{equation}
at layer $N$. At the remaining layers, if the cancellation at layer $N$ is correct, we get $\acute{P}_n(\rp_{nn}) = Q \left( \sqrt{ \frac{ 2\rp_{nn}^2 }{ \sigma^{2} } } \right)$. Otherwise, we have
\begin{equation}\label{eq:cancellation}
\bar{y}_{n}\!-\! \rp_{nN}^{}\hat{x}_{N}^{} = \rp_{nn}x_{n} + \rp_{nN}^{}(x_{N}^{} - \hat{x}_{N}^{}) + w_n.
\end{equation}
Noting that $x_{N}^{}\!-\!\hat{x}_{N}^{}\!=\!\pm 2$, the variance of interference plus noise is $\sigma^{2}\!+\!4$, and hence the BER becomes $\acute{P}_n(\rp_{nn}) = Q \left(  \sqrt{ \frac{ 2\rp_{nn}^2 }{ \sigma^{2} + 4 } } \right) $. Thus, the resultant BER for detecting $x_n$ can be written as
\begin{equation}\label{eq:berPSIC}
     \acute{P}_n(\rp_{nn})=  Q \left( \sqrt{ \frac{ 2\rp_{nn}^2 }{ \sigma^{2} } } \right) \left(1-\acute{P}_N^{}(\rp_{NN}^{})\right) + Q \left( \sqrt{ \frac{ 2\rp_{nn}^2 }{ \sigma^{2} + 4 } } \right) \acute{P}_N^{}(\rp_{NN}^{}).
\end{equation}
The BER for N/C with regular QRD at layer $n$ ($n\!=\!N\!-\!1,\cdots,1$) is \cite{choi2006nulling}
\begin{align}\label{eq:berSIC}
P_n(r_{nn}) & = \sum_{\psi_{n+1}\in D_{n+1}} P_n(\err|r_{nn},\psi_{n+1})P_{n+1}(\psi_{n+1}) \\
    P_n(\err|r_{nn},\psi_{n+1}) & = Q \left( \sqrt{ \frac{ 2r_{nn}^2 }{ \sigma^{2} + 4\psi_{n+1}\psi_{n+1}^{T} } } \right),
\end{align}
where $P_{n+1}(\psi_{n+1})$ can be computed recursively, and $\psi_n$ is an instance of $D_n$, the set of all possible error patterns leading to layer $n$, which are represented as binary vectors with $1$ in the place of incorrect layer detection:
\begin{equation}\label{eq:Dn}
     D_n = \{[\underbrace{0 0 \cdots 0}_{N-n+1}],[\underbrace{0 0 \cdots 1}_{N-n+1}],\cdots[\underbrace{1 1 \cdots 1}_{N-n+1}] \}.
\end{equation}
Note that with WRD, we have a smaller set $\acute{D}_n = \{[0 0 \cdots 0],[0 0 \cdots 1] \} \subset D_n$, where $\abs{\acute{D}_n} = 2 < \abs{D_n} = 2^{N-n+1}$. Therefore, error propagation is largely reduced.

However, having fewer terms in the BER formula does not mean a better BER performance. The average BER at layer $n$ is obtained by taking the expectation over $r_{nn}^2$ and $\rp_{nn}^2$. Since $\rp_{nn}^2$ has smaller values than $r_{nn}^2$ at layers $1\!\leq\!n\!\leq\!N\!-\!2$, and since $Q(\cdot)$ is a monotonically decreasing function, PN/C will result in performance degradation. But layer $N$ has the worst performance, despite not being affected by noise coloring, since more errors can happen at this layer due to low array gain, and they get propagated to higher layers. Therefore, the performance of both N/C and PN/C detectors will be dominated by $\acute{P}_N^{}(\rp_{NN}^{})=P_N^{}(r_{NN}^{})$, and all computational savings in the proposed PN/C detector will come at a negligible cost.

In equation form, the function $G(d,\gamma)$ provides the average BER over a $d$-fold diversity Rayleigh fading channel with mean branch SNR $\gamma$:
\begin{equation}\label{eq:ber_avg_2}
    G \left( d, \gamma\right) = \left[ \frac{1}{2} \left( 1-\mu\right)\right]^d \sum_{k=0}^{d-1} \left( \begin{array}{c} d-1+k \\ k \end{array} \right)\left[ \frac{1}{2} \left( 1+\mu\right) \right]^k,
\end{equation}
where $\mu=\sqrt{\gamma/(1+\gamma)}$. The average BER with PN/C at layers $1\!\leq\!n\!\leq\!N\!-\!1$ is thus expressed as
\begin{equation}\label{eq:ber_avg_1}
     \acute{P}_n =  G \left( 2, \frac{1}{\sigma^{2}} \right) (1-\acute{P}_N) + G \left( 2, \frac{1}{\sigma^{2} + 4} \right) \acute{P}_N,
\end{equation}
where layers $1\!\leq\!n\!\leq\!N\!-\!1$ only provide a $2$-fold diversity due to puncturing. Similarly, the average BERs at layer $n<N$ for N/C can be obtained by replacing $P_n(\err|r_{nn},\psi_{n+1})$ in equation \eqref{eq:berSIC} by its average over $r_{nn}$, $P_n(\err|\psi_{n+1})$, where
\begin{equation}\label{eq:ber_avg_3}
    P_n(\err|\psi_{n+1}) =  \mathsf{E}[P_n(\err|r_{nn},\psi_{n+1})] = G \left( N-n+1, \frac{1}{\sigma^{2} + 4\psi_{n+1}\psi_{n+1}^{T}} \right),
\end{equation}
with the numerator in $\gamma$ being 1 because we assume normalization. We have
\begin{equation}\label{eq:ber_avg_4}
    \acute{P}_N^{}(\rp_{NN}^{}) = P_N^{}(r_{NN}^{}) =  G(1,1/\sigma^{2}) = \frac{1}{2} \left( 1 - \sqrt{\frac{1/\sigma^{2}}{1+1/\sigma^{2}}} \right).
\end{equation}
Note that these equations can be extended to an arbitrary constellation size by expressing the function $G(d,\gamma)$ as $G(d,\gamma,\abs{\mathcal{M}})$ as shown in Appendix \ref{FirstAppendix}.

%
\subsection{Punctured Chase Detector (PCD)}\label{sec:berCD}
To capture the performance of the PCD, we follow a probabilistic approach similar to that in \cite{radji2009interference}. Denote by $\hat{\mbf{x}}^{\ML}$, $\hat{\mbf{x}}^{\CD}$, and $\hat{\mbf{x}}^{\PCD}$, the vector outputs, and by $P^{\ML}$, $P^{\CD}$, and $P^{\PCD}$, the vector error rates, of the ML detector, the CD, and the PCD, respectively. We start by the case of a regular CD; we have
\begin{equation}\label{eq:prob1}
    P^{\CD} = \mathsf{E}_{\mbf{R},\mbf{x}} \bigg[\Prb\left(\hat{\mbf{x}}^{\CD}\neq\mbf{x}~|~\mbf{R},\mbf{x}\right)\bigg].
\end{equation}
For clarity of presentation, we drop the expectation operator in what follows. Clearly, $P^{\ML}<P^{\CD}$ is a lower bound; we seek a tight upper bound. We further have
\begin{align}
  P^{\CD} & = \Prb(\hat{\mbf{x}}^{\CD}\neq\mbf{x}~|~\mbf{R},\mbf{x}) \notag\\
  & = \Prb(\hat{\mbf{x}}^{\CD}\neq\mbf{x}~|~\mbf{R},\mbf{x},\hat{\mbf{x}}^{\ML}\neq\mbf{x})\Prb(\hat{\mbf{x}}^{\ML}\neq\mbf{x}~|~\mbf{R},\mbf{x}) \notag\\
  & + \Prb(\hat{\mbf{x}}^{\CD}\neq\mbf{x}~|~\mbf{R},\mbf{x},\hat{\mbf{x}}^{\ML}=\mbf{x})\Prb(\hat{\mbf{x}}^{\ML}=\mbf{x}~|~\mbf{R},\mbf{x}).\label{eq:prob2}
\end{align}
The first term in equation \eqref{eq:prob2} is upper bounded by $P^{\ML}=\Prb(\hat{\mbf{x}}^{\ML}\neq\mbf{x}~|~\mbf{R},\mbf{x})$. To simplify the second term, we expand
\begin{align}\label{eq:prob3}
  \Prb(\hat{\mbf{x}}^{\CD}\neq\mbf{x}~|~\mbf{R},\mbf{x},\hat{\mbf{x}}^{\ML}=\mbf{x})
  & = \Prb(~\hat{\mbf{x}}^{\CD}\neq\mbf{x}~|~\mbf{R},\mbf{x},\hat{\mbf{x}}^{\ML}=\mbf{x},\mbf{x}\in \mathcal{S}(\mbf{\tilde{y}},\mbf{R})~) \\
  & \times \Prb(\mbf{x}\in \mathcal{S}(\mbf{\tilde{y}},\mbf{R})~|~\mbf{R},\mbf{x},\hat{\mbf{x}}^{\ML}=\mbf{x}) \\
  & + \Prb(~\hat{\mbf{x}}^{\CD}\neq\mbf{x}~|~\mbf{R},\mbf{x},\hat{\mbf{x}}^{\ML}=\mbf{x},\mbf{x}\not\in \mathcal{S}(\mbf{\tilde{y}},\mbf{R})~) \\
  & \times \Prb(\mbf{x}\not\in \mathcal{S}(\mbf{\tilde{y}},\mbf{R})~|~\mbf{R},\mbf{x},\hat{\mbf{x}}^{\ML}=\mbf{x}).
\end{align}
We can note that
\begin{align}\label{eq:prob4}
& \Prb(~\hat{\mbf{x}}^{\CD}\neq\mbf{x}~|~\mbf{R},\mbf{x},\hat{\mbf{x}}^{\ML}=\mbf{x},\mbf{x} \in \mathcal{S}(\mbf{\tilde{y}},\mbf{R})~) \triangleq P^{\A} = 0,~\text{and} \\
& \Prb(~\hat{\mbf{x}}^{\CD}\neq\mbf{x}~|~\mbf{R},\mbf{x},\hat{\mbf{x}}^{\ML}=\mbf{x},\mbf{x} \not\in \mathcal{S}(\mbf{\tilde{y}},\mbf{R})~) = 1.
\end{align}
Hence, substituting back in equation \eqref{eq:prob3}, we get
\begin{equation}\label{eq:prob5}
  \Prb(\hat{\mbf{x}}^{\CD}\neq\mbf{x}~|~\mbf{R},\mbf{x},\hat{\mbf{x}}^{\ML}=\mbf{x})
  = \Prb(\mbf{x}\not\in \mathcal{S}(\mbf{\tilde{y}},\mbf{R})~|~\mbf{R},\mbf{x},\hat{\mbf{x}}^{\ML}=\mbf{x}) \triangleq P^{\B}.
\end{equation}
Substituting back in~\eqref{eq:prob2}, the second term is upper bounded by $P^{\B}$, which effectively is equivalent to the probability that the generated list does not contain the true vector. But since we exhaustively search over all possible values of $\hat{x}_N^{}$ in $\mathcal{M}$, this probability will in effect be the probability of error in SIC on upper layers. Therefore, we have
\begin{equation}\label{eq:boundsCD}
P^{\ML}<P^{\CD}<P^{\ML}+P^{\B}.
\end{equation}

\begin{figure}[t]
\centering
\includegraphics[width=0.55\linewidth]{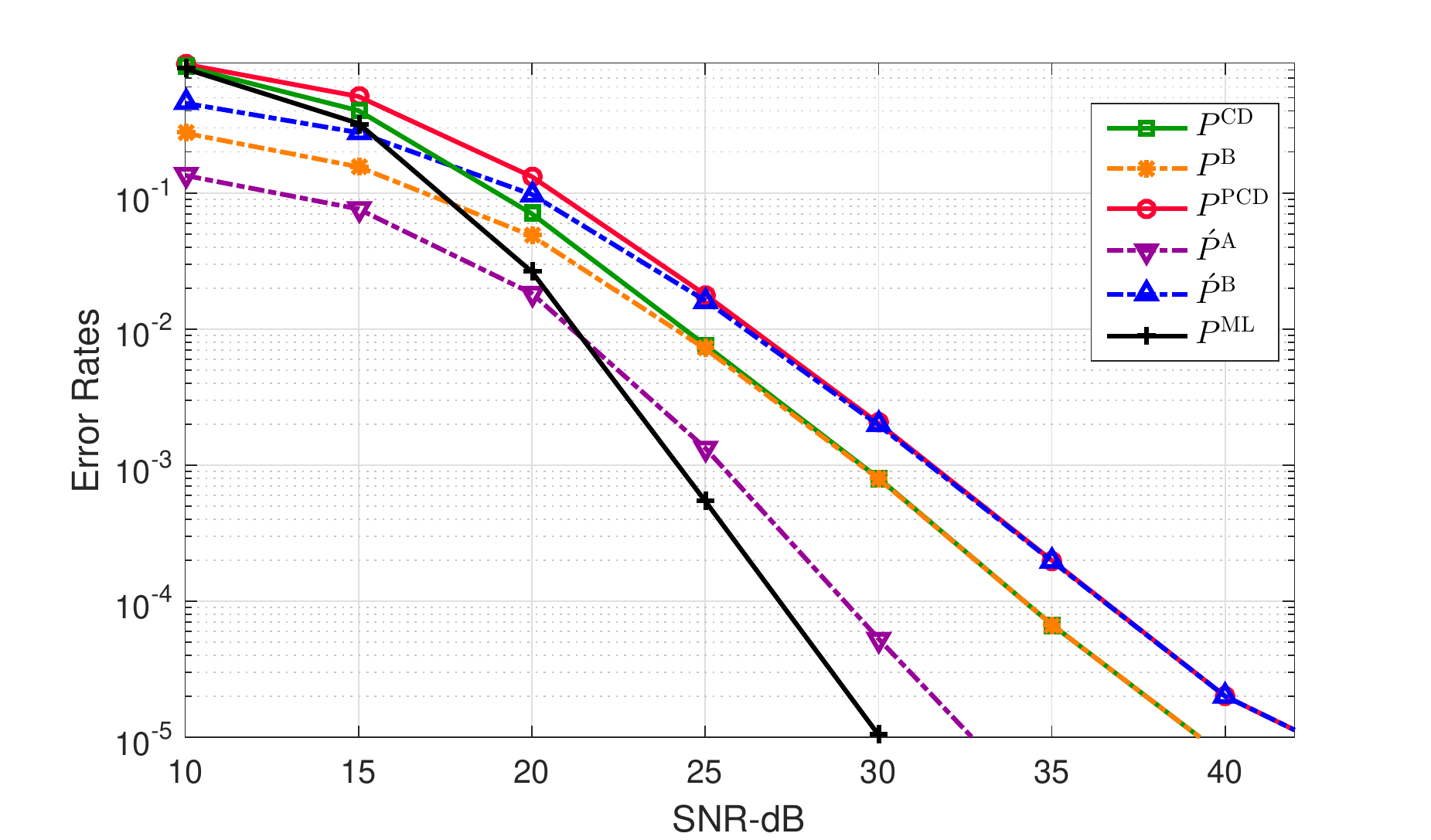}\vspace{-0.175in}
\caption{Empirical simulations of the error probabilities of interest with chase detection when $\mathcal{M}$ is $16$-QAM.}
\label{perffig:a}
\end{figure}

Similarly, we can derive the bounds for $P^{\PCD}$ by expanding $\Prb(\hat{\mbf{x}}^{\PCD}\neq\mbf{x}~|~\Rp,\mbf{x})$ and substituting $\mathcal{S}(\mbf{\tilde{y}},\mbf{R})$ with $\mathcal{P}(\mbf{\bar{y}},\Rp)$. However, the modified $P^{\A}$ and $P^{\B}$, $\acute{P}^{\A}$ and $\acute{P}^{\B}$, will evaluate differently:
\begin{align}\label{eq:prob6}
& \acute{P}^{\A}=\Prb(~\hat{\mbf{x}}^{\PCD}\!\neq\!\mbf{x}~|~\Rp,\mbf{x},\hat{\mbf{x}}^{\ML}\!=\!\mbf{x},\mbf{x} \!\in \!\mathcal{P}(\mbf{\bar{y}},\Rp)~) \neq 0 \\
& \acute{P}^{\B}= \Prb(\mbf{x}\not\in \mathcal{P}(\mbf{\bar{y}},\Rp)~|~\Rp,\mbf{x},\hat{\mbf{x}}^{\ML}=\mbf{x}) > P^{\B}.
\end{align}
Note that $\acute{P}^{\A}$ is not zero, because even if $\hat{\mbf{x}}^{\ML}$ is within the generated list, the modified distance metric might not select it as the HO vector. Therefore, the bounds for the PCD are
\begin{equation}\label{eq:boundsPCD}
P^{\ML}<P^{\PCD}<P^{\ML}+\acute{P}^{\B}+\xi,
\end{equation}
where $\xi=\acute{P}^{\A}(1-\acute{P}^{\B})(1-P^{\ML})$. As shown in Fig.~\ref{perffig:a}, the dominant factors that affect the performances of $P^{\CD}$ and $P^{\PCD}$ at high $\mathsf{SNR}$ are $P^{\B}$ and $\acute{P}^{\B}$, respectively. Thus, $\acute{P}^{\A}$, and hence $\xi$, can be safely neglected. Nevertheless, since SIC over a punctured channel matrix is more prone to errors, we have $\acute{P}^{\B} > P^{\B}$, and accordingly $P^{\PCD} > P^{\CD}$. But layers with smallest degrees of freedom (equal to 4) dominate the BER performance, and they exist in both the CD and the PCD. Therefore, the performance gap will be in the form of a shift, and there is no loss in diversity gain.

An alternative approximate approach to study the performance of the CD starting from N/C exists \cite{waters2008chase}, where the gain of generating a list by considering more candidate symbols at layer $N$ can be seen as an effective $\mathsf{SNR}$ gain at that layer. The gain factor is given by
\begin{equation}\label{eq:SNRgain}
\Gamma = \frac{\delta_{Ls}}{\delta_N^{}},
\end{equation}
where $\delta_N^{}$ is the distance between the transmitted symbol and the nearest decision boundary in $\mathcal{M}$, and $\delta_{Ls}$ is the distance to the nearest decision boundary of the list (an error occurs when $x_N^{}$ is not in the list). Since in our case the list is the entirety of $\mathcal{M}$, the $\mathsf{SNR}$ gain is $\Gamma=\infty$. In other words, no error is propagated from layer $N$, which was the limiting layer in N/C. Therefore, the BER of CD and PCD can be obtained by summing the combinations of BERs on all layers $n<N$, which are computed using equations \eqref{eq:berSIC} and \eqref{eq:berPSIC}, respectively, while setting $\acute{P}_N^{}(\rp_{NN}^{}) = P_N^{}(r_{NN}^{}) = 0$ in these equations. Hence, we have
\begin{equation}\label{eq:ber_PCD}
     P^{\PCD} =  (N-1) \left[ G \left( 2, \frac{1}{\sigma^{2}} \right) \right].
\end{equation}

\begin{figure*}[t]
  \centering
  \subfloat[$4\times4$ MIMO]{\label{theofig:a} \includegraphics[width=0.48\linewidth]{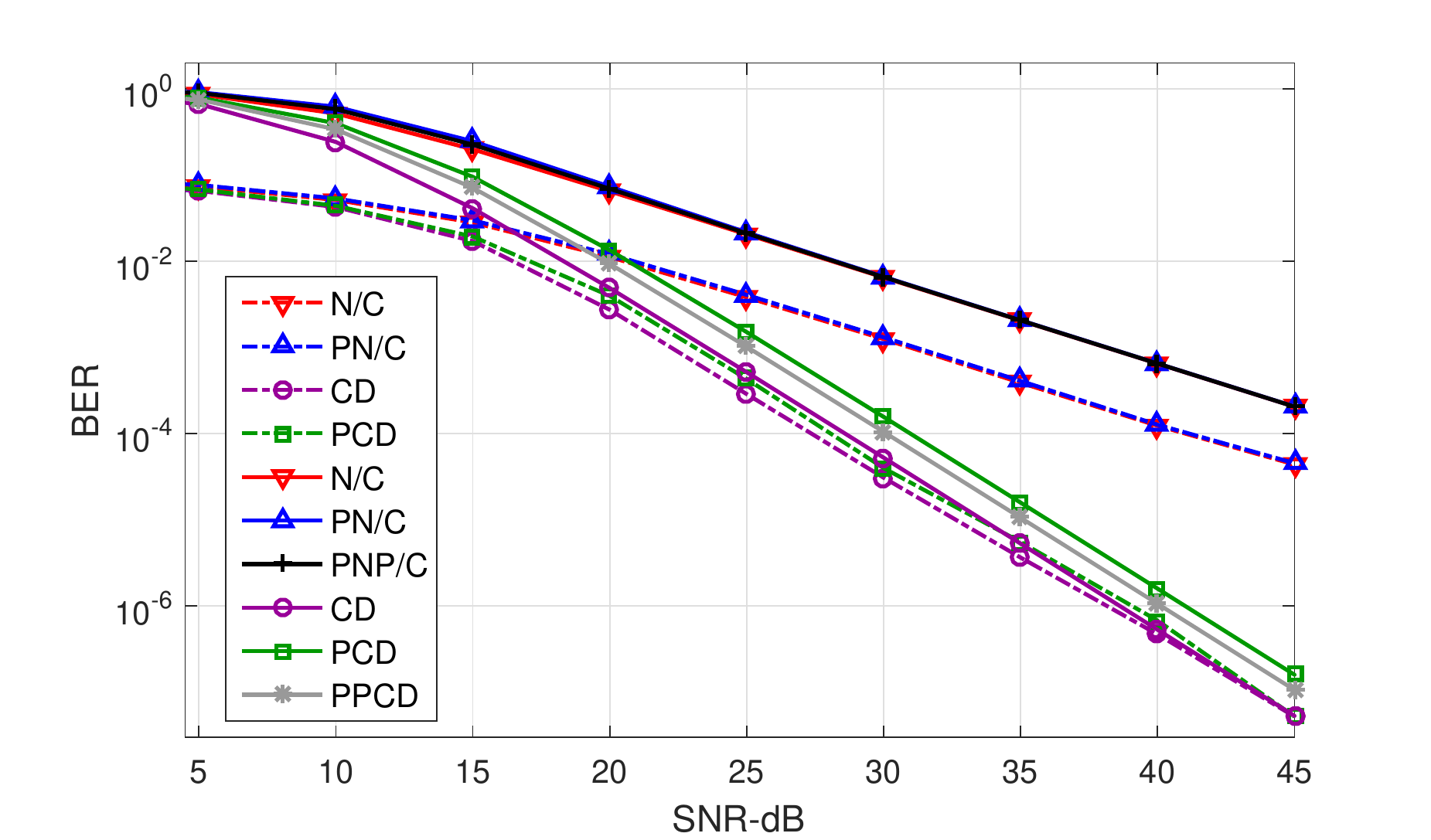}}
  \hfill
  \subfloat[$16\times16$ MIMO]{\label{theofig:b} \includegraphics[width=0.48\linewidth]{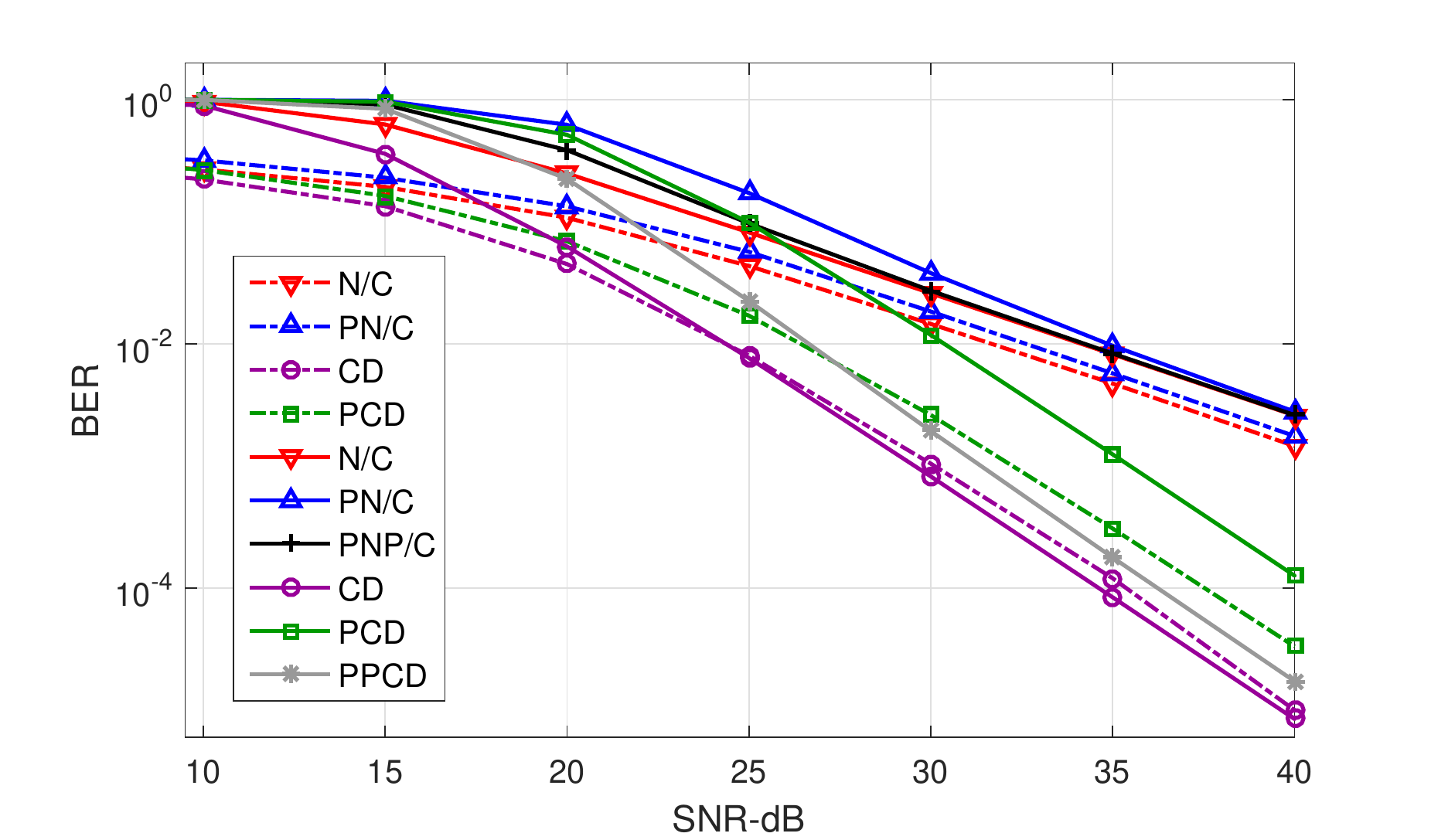}}

  \caption{Theoretical BERs (solid lines) vs simulated BERs (dotted lines) of various CDs and N/C detectors with $16$-QAM.}
  \label{f:theoretical}
\end{figure*}

Figure~\ref{f:theoretical} shows the theoretical and simulated BERs with $16$-QAM, for $4\times4$ and $16\times16$ MIMO systems. Note that the CD and PCD theoretical curves are not true upper bounds on error performance, but rather close approximations \cite{waters2008chase}. The gap between punctured and unpunctured schemes increases with the number of antennas. However, this gap can be reduced by employing partial puncturing, at the expense of less computational savings. For example, the partially-punctured N/C (PPN/C) detector and the partially-punctured CD (PPCD) in Fig. \ref{f:theoretical}, correspond to the case where the entries of the third column above the diagonal are not zeroed-out during puncturing. In the remainder of this paper, we always consider full puncturing that results in maximal complexity reduction.

%
\subsection{Symbol-Based Sub-Space Detector (SSSD)}
\label{sec:berSSsb}

Denote by $P^{\SLORD}$ and $P^{\SSSD}$ the vector error rates, of SSSD and SLORD, respectively. We start by the case of SLORD. Noting that the performance can be captured from one partition, say $t\!=\!1$ (Fig. \ref{f:matrices}-a), we investigate:
\begin{align}\label{eq:prob22}
  \Prb(\hat{x}^{\SLORD}_N\neq x_N^{}~|~\mbf{R},\mbf{x}) & = \Prb(\hat{x}^{\SLORD}_N\neq x_N^{}~|~\mbf{R},\mbf{x},\hat{\mbf{x}}^{\ML}\neq\mbf{x})\Prb(\hat{\mbf{x}}^{\ML}\neq\mbf{x}~|~\mbf{R},\mbf{x}) \\
  & + \Prb(\hat{x}^{\SLORD}_N\neq x_N^{}~|~\mbf{R},\mbf{x},\hat{\mbf{x}}^{\ML}=\mbf{x})\Prb(\hat{\mbf{x}}^{\ML}=\mbf{x}~|~\mbf{R},\mbf{x})
\end{align}
\begin{align}\label{eq:prob32}
  \Prb(\hat{x}^{\SLORD}_N\neq x_N^{}~|~\mbf{R},\mbf{x},\hat{\mbf{x}}^{\ML}=\mbf{x}) &  = \Prb(~\hat{x}^{\SLORD}_N\neq x_N^{}~|~\mbf{R},\mbf{x},\hat{\mbf{x}}^{\ML}=\mbf{x},\mbf{x}\in \mathcal{S}(\mbf{\tilde{y}},\mbf{R})~) & \\
  & \times \Prb(\mbf{x}\in \mathcal{S}(\mbf{\tilde{y}},\mbf{R})~|~\mbf{R},\mbf{x},\hat{\mbf{x}}^{\ML}=\mbf{x}) \\
  & + \Prb(~\hat{x}^{\SLORD}_N\neq x_N^{}~|~\mbf{R},\mbf{x},\hat{\mbf{x}}^{\ML}=\mbf{x},\mbf{x}\not\in \mathcal{S}(\mbf{\tilde{y}},\mbf{R})~) \\
  & \times \Prb(\mbf{x}\not\in \mathcal{S}(\mbf{\tilde{y}},\mbf{R})~|~\mbf{R},\mbf{x},\hat{\mbf{x}}^{\ML}=\mbf{x}).
\end{align}
We define:
\begin{align}\label{eq:prob42}
& \Prb(~\hat{x}^{\SLORD}_N\neq x_N^{}~|~\mbf{R},\mbf{x},\hat{\mbf{x}}^{\ML}=\mbf{x},\mbf{x} \in \mathcal{S}(\mbf{\tilde{y}},\mbf{R})~) ~\triangleq~ P^{\D} = 0 \\
& \Prb(~\hat{x}^{\SLORD}_N\neq x_N^{}~|~\mbf{R},\mbf{x},\hat{\mbf{x}}^{\ML}=\mbf{x},\mbf{x} \not\in \mathcal{S}(\mbf{\tilde{y}},\mbf{R})~) ~\triangleq~ P^{\C} \neq 1\\
& \Prb(\mbf{x}\not\in \mathcal{S}(\mbf{\tilde{y}},\mbf{R})~|~\mbf{R},\mbf{x},\hat{\mbf{x}}^{\ML}=\mbf{x}) = P^{\B},
\end{align}
and substitute back in equation \eqref{eq:prob32} to get
\begin{equation}\label{eq:prob52}
  \Prb(\hat{x}^{\SLORD}_N\neq x_N^{}~|~\mbf{R},\mbf{x},\hat{\mbf{x}}^{\ML}=\mbf{x}) = P^{\C}P^{\B}.
\end{equation}
Following the same argument as in Sec.\ref{sec:berCD} we have.
\begin{align}\label{eq:boundsSLORD}
& P^{\ML}<P^{\SLORD}<P^{\ML}+P^{\C}P^{\B}(1-P^{\ML})\\
& P^{\ML}<P^{\SLORD}<P^{\ML}+P^{\C}P^{\B}
\end{align}

Similarly, we can derive the bounds for $P^{\SSSD}$ by expanding $\Prb(\hat{x}^{\SSSD}_N\neq x_N^{}~|~\Rp,\mbf{x})$. We define
\begin{align}\label{eq:prob43}
& \Prb(~\hat{x}^{\SSSD}_N\neq x_N^{}~|~\Rp,\mbf{x},\hat{\mbf{x}}^{\ML}=\mbf{x},\mbf{x}\!\in \mathcal{P}(\mbf{\bar{y}},\Rp)~) = \acute{P}^{\D} \neq 0 \\
& \Prb(~\hat{x}^{\SSSD}_N\neq x_N^{}~|~\Rp,\mbf{x},\hat{\mbf{x}}^{\ML}=\mbf{x},\mbf{x} \not\in \mathcal{P}(\mbf{\bar{y}},\Rp)~) = \acute{P}^{\C} \neq 1 \\
& \Prb(\mbf{x}\not\in \mathcal{P}(\mbf{\bar{y}},\Rp)~|~\Rp,\mbf{x},\hat{\mbf{x}}^{\ML}=\mbf{x}) = \acute{P}^{\B},
\end{align}
where following the same procedure we get
\begin{align}\label{eq:boundsSLORD2}
& P^{\ML}<P^{\SSSD}<P^{\ML}+(~\acute{P}^{\C}\acute{P}^{\B}+\acute{P}^{\D}(1-\acute{P}^{\B})~)(1-P^{\ML}) \\
& P^{\ML}<P^{\SSSD}<P^{\ML}+\acute{P}^{\C}\acute{P}^{\B}+\acute{P}^{\D}.
\end{align}

\begin{figure}[t]
\centering
\includegraphics[width=0.55\linewidth]{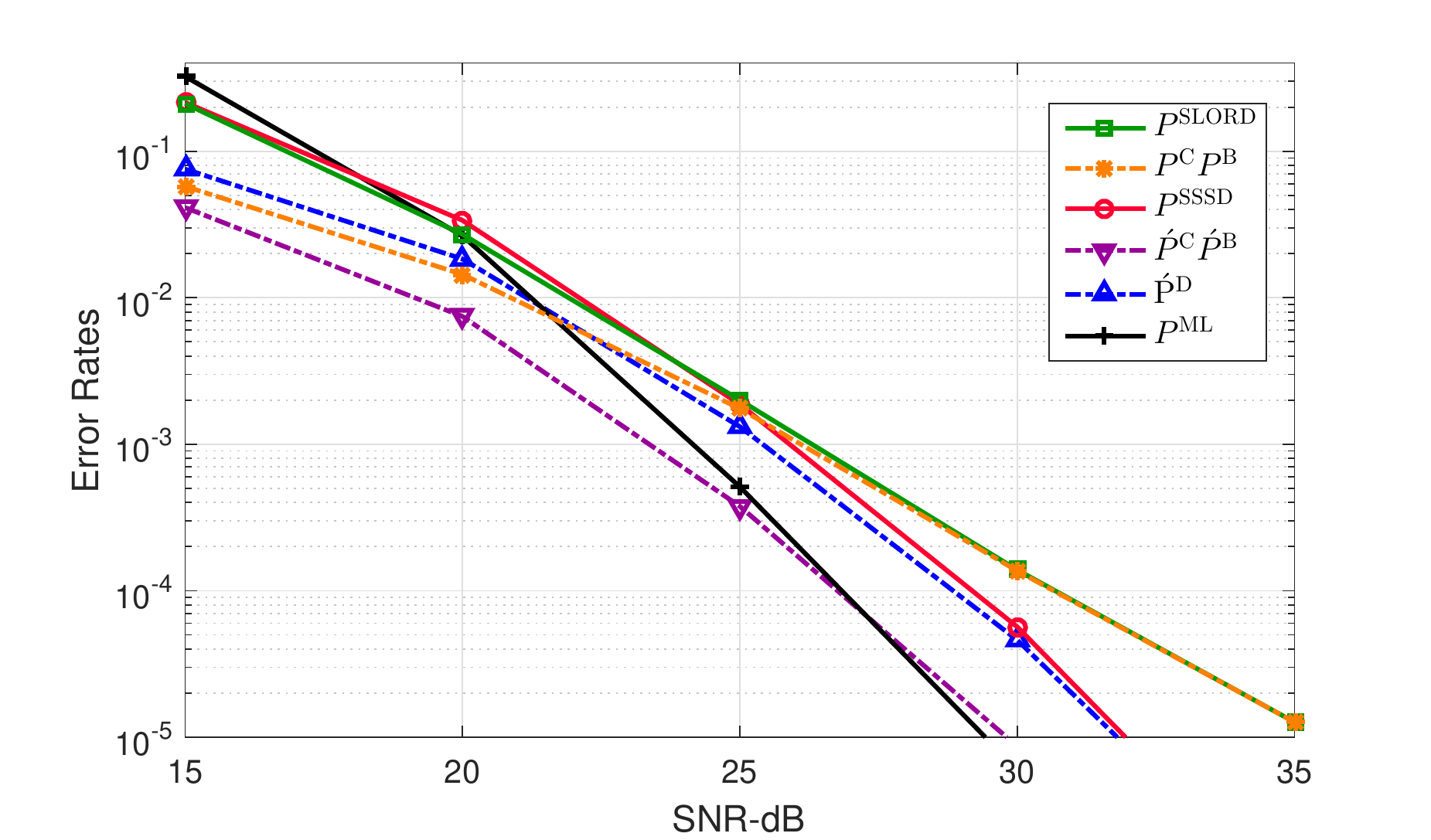}\vspace{-0.175in}
\caption{Empirical simulations of the error probabilities of interest with chase detection when $\mathcal{M}$ is $16$-QAM.}
\label{perffig:b}
\end{figure}

Therefore, the dominant factor that affects $P^{\SLORD}$ at high $\mathsf{SNR}$ is $P^{\C}P^{\B}$, and the factors that affect $P^{\SSSD}$ are $\acute{P}^{\C}\acute{P}^{\B}$ and $\acute{P}^{\D}$ (we can show that $\acute{P}^{\D}$ is identical to $\acute{P}^{\A}$ from equation \eqref{eq:prob6}). Note that $P^{\C}$ can be expressed as the probability that an error occurs at layer $N$ in the CD, given that an error occurred at the upper layers, and $\acute{P}^{\C}$ is similarly defined in the case of a PCD. Since the PCD does not propagate errors at upper layers, the distance metric \eqref{eq:xwr} is not severely distorted, and $\hat{x}_N^{}$ can still be recovered. We thus have $\acute{P}^{\C}\ll P^{\C}$. Moreover, as Fig.\ref{perffig:b} shows, $\acute{P}^{\D}$ is the limiting term for $P^{\SSSD}$ at high $\mathsf{SNR}$ because $\acute{P}^{\C}\acute{P}^{\B}\ll \acute{P}^{\D}$. Hence, although $P^{\B} < \acute{P}^{\B}$, we still have $P^{\SSSD} < P^{\SLORD}$, which is a gain caused by puncturing. This is in accordance with the conclusion in Sec.~\ref{sec:diversitySSSD}.

%
\section{Soft Output Detection}
\label{sec:soft_output}

To generate the log-likelihood ratios (LLRs) with SSSD, the $N$ streams should be decoupled in $N$ steps, where in each step $t \in \{1,\cdots,N\}$ the LLRs for the bits corresponding to symbol $x_n$ ($n=t$) are calculated. Hence, for each bit, we compute
\begin{equation}\label{eq:LLR_SSSD}
  \lambda_{n,k,t}^{\SSSD}\!=\!\frac{1}{\sigma^{2}} \left(\min_{\mbf{x} \in \mathcal{P}^{n,k,t,0}}{\norm{\mbf{\bar{y}}^{(t)}\!-\!\Rp^{(t)}\mbf{x}}^{2}}\!-\!\min_{\mbf{x} \in \mathcal{P}^{n,k,t,1}}{\norm{\mbf{\bar{y}}^{(t)}\!-\!\Rp^{(t)}\mbf{x}}^{2}} \right)
\end{equation}
for $t\!=\!1,\!\cdots\!,N$ and $k\!=\!1,\!\cdots\!,\log_2(\abs{\mathcal{M}})$, where the sets $\mathcal{P}^{n,k,t,0}\!\triangleq\!\{\mbf{x} \in \mathcal{P}(\mbf{\bar{y}^{(t)}},\Rp^{(t)}): b_{n,k}\!=\!0\}$ and $\mathcal{P}^{n,k,t,1}\!\triangleq\!\{\mbf{x} \in \mathcal{P}(\mbf{\bar{y}^{(t)}},\Rp^{(t)}): b_{n,k}\!=\!1\}$ correspond to subsets of symbol vectors in $\mathcal{P}(\mbf{\bar{y}^{(t)}},\Rp^{(t)})$, having in the corresponding $\nth{k}$ bit of the $\nth{n}$ symbol a value of $0$ and $1$, respectively. Note that these distance metrics can be expanded as in equation \eqref{eq:xwr}. Similarly, we can define the LLRs for SLORD as
\begin{equation}\label{eq:LLRSLORD}
  \lambda_{n,k,t}^{\SLORD}\!=\!\frac{1}{\sigma^{2}} \left(\min_{\mbf{x} \in \mathcal{S}^{n,k,t,0}}{\norm{\mbf{\tilde{y}}^{(t)}\!-\!\mbf{R}^{(t)}\mbf{x}}^{2}}\!-\!\min_{\mbf{x} \in \mathcal{S}^{n,k,t,1}}{\norm{\mbf{\tilde{y}}^{(t)}\!-\!\mbf{R}^{(t)}\mbf{x}}^{2}} \right),
\end{equation}
where $\mathcal{S}^{n,k,t,0}\!\triangleq\!\{\mbf{x} \in \mathcal{S}(\mbf{\tilde{y}^{(t)}},\mbf{R}^{(t)}): b_{n,k}\!=\!0\}$ and $\mathcal{S}^{n,k,t,1}\!\triangleq\!\{\mbf{x} \in \mathcal{S}(\mbf{\tilde{y}^{(t)}},\mbf{R}^{(t)}): b_{n,k}\!=\!1\}$.
\begin{figure}[t]
\centering
\includegraphics[scale=0.455]{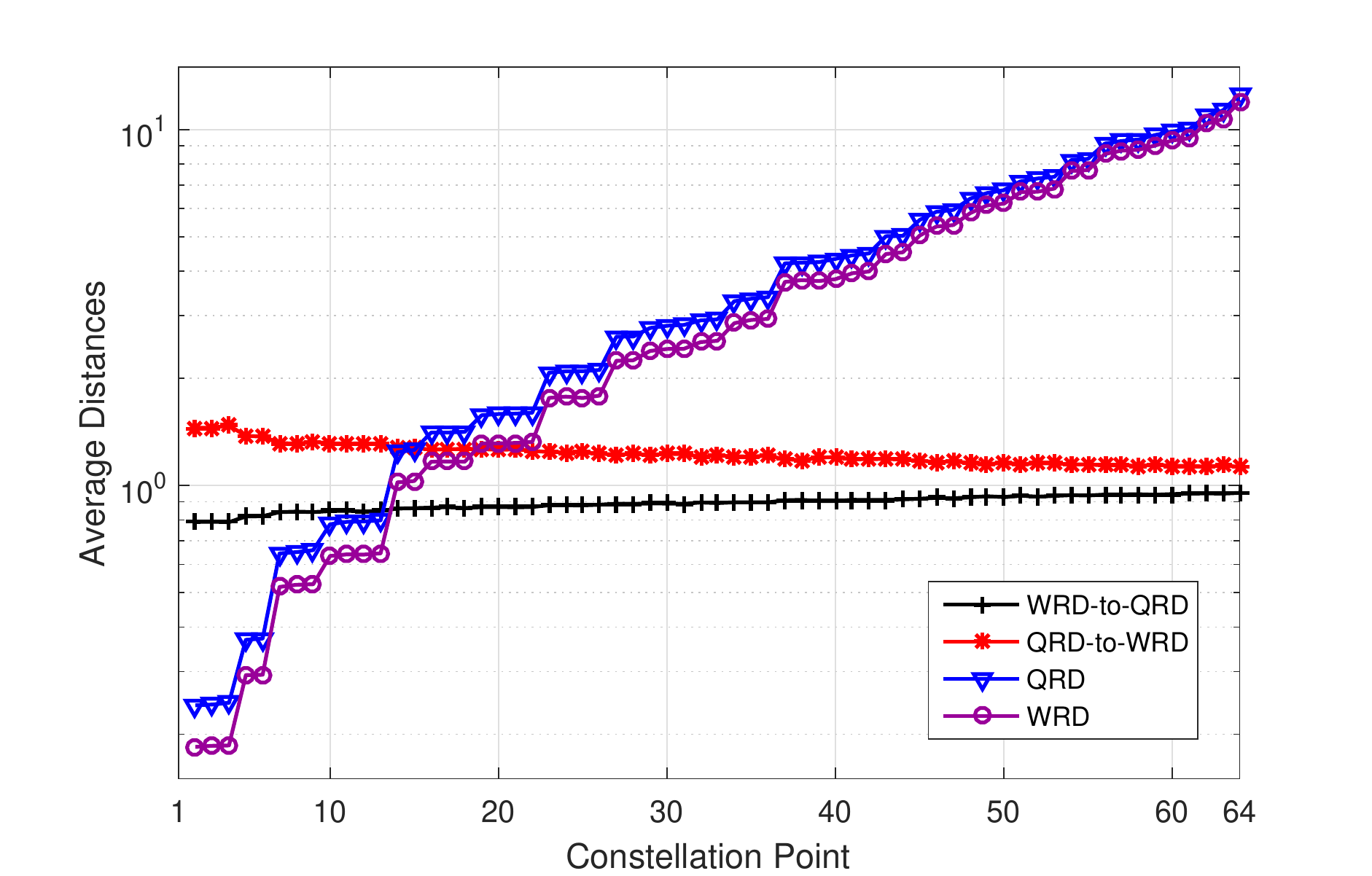}\vspace{-0.15in}
\caption{Comparison of QRD-based distance metrics ($\norm{\mbf{Q}^*(\mbf{y} - \mbf{H}\mbf{x})}^{2}$) and WRD-based distance metrics ($\norm{\mbf{W}^*(\mbf{y} - \mbf{H}\mbf{x})}^{2}$) versus symbols from the constellation at the root layer. A noiseless $4\times4$ MIMO system with $64$-QAM is assumed. The symbols on the x-axis are sorted in increasing order of distance metrics (constellation point 1 with a zero distance corresponds to the true transmitted symbol vector). Both absolute distances as well as distance ratios are shown.}
\label{f:scalings}
\end{figure}
Using the results of Sec.~\ref{sec:berSSsb}, we deduce that SO SSSD outperforms SO SLORD. Moreover, note that with CD and PCD, only the layer of interest is exhaustively searched, which provides the required distance metrics to compute the LLRs on this layer only. Repeating the same process on all layers, the resultant detectors would be SO SLORD and SO SSSD, respectively:
\begin{equation}\label{eq:equality}
  \lambda_{n,k,t}^{\PCD}\!=\!\lambda_{n,k,t}^{\SSSD}, \ \ \ \lambda_{n,k,t}^{\CD}\!=\!\lambda_{n,k,t}^{\SLORD}.
\end{equation}

With SO SSD and LORD, tighter LLRs can be computed by tracking global distances rather than per stream distances. This requires an extra processing overhead as follows:
\begin{equation}\label{eq:LLRSSD}
  \lambda_{n,k}^{\SSD} = \frac{1}{\sigma^{2}} \bigg[ \min_t \left( \min_{\mbf{x} \in \mathcal{P}^{n,k,t,0}}{\norm{\mbf{\bar{y}}^{(t)} - \Rp^{(t)}\mbf{x}}^{2}} \right) - \min_t \left( \min_{\mbf{x} \in \mathcal{P}^{n,k,t,1}}{\norm{\mbf{\bar{y}}^{(t)} - \Rp^{(t)}\mbf{x}}^{2}} \right) \bigg]
\end{equation}
\begin{equation}\label{eq:LLRLORD}
  \lambda_{n,k}^{\LORD} = \frac{1}{\sigma^{2}} \bigg[ \min_t \left( \min_{\mbf{x} \in \mathcal{S}^{n,k,t,0}}{\norm{\mbf{\tilde{y}}^{(t)} - \mbf{R}^{(t)}\mbf{x}}^{2}} \right) - \min_t \left( \min_{\mbf{x} \in \mathcal{S}^{n,k,t,1}}{\norm{\mbf{\tilde{y}}^{(t)} - \mbf{R}^{(t)}\mbf{x}}^{2}} \right) \bigg].
\end{equation}
But since distance metrics that are computed from different channel decompositions in SSD are independent, SO SSD will not achieve a better performance than SO SSSD.

Finally, the performance of SO detectors is sensitive to factors that are not captured in the HO analysis. For example, computing distances via WRD as $\norm{\mbf{W}^*(\mbf{y} - \mbf{H}\mbf{x})}^{2}$, instead of $\norm{\mbf{Q}^*(\mbf{y} - \mbf{H}\mbf{x})}^{2}$ is subject to downscaling, as Fig. \ref{f:scalings} shows. The scaling effect is higher with smaller distances, that will end up as ML or counter-ML distances to be used in LLR computations. Hence, the LLRs get scaled accordingly. This results in less confidence in LLR outputs, which is beneficial since LORD schemes produce overconfident LLRs. Furthermore, Fig. \ref{f:scalings} shows that on average, the order of distance metrics is retained under puncturing. However, at specific instances, the order gets distorted with a probability greater than $\acute{P}^{\A}$.

\begin{figure*}[t]
  \centering
  \subfloat[]{\label{archfig:a} \includegraphics[width=0.45\linewidth]{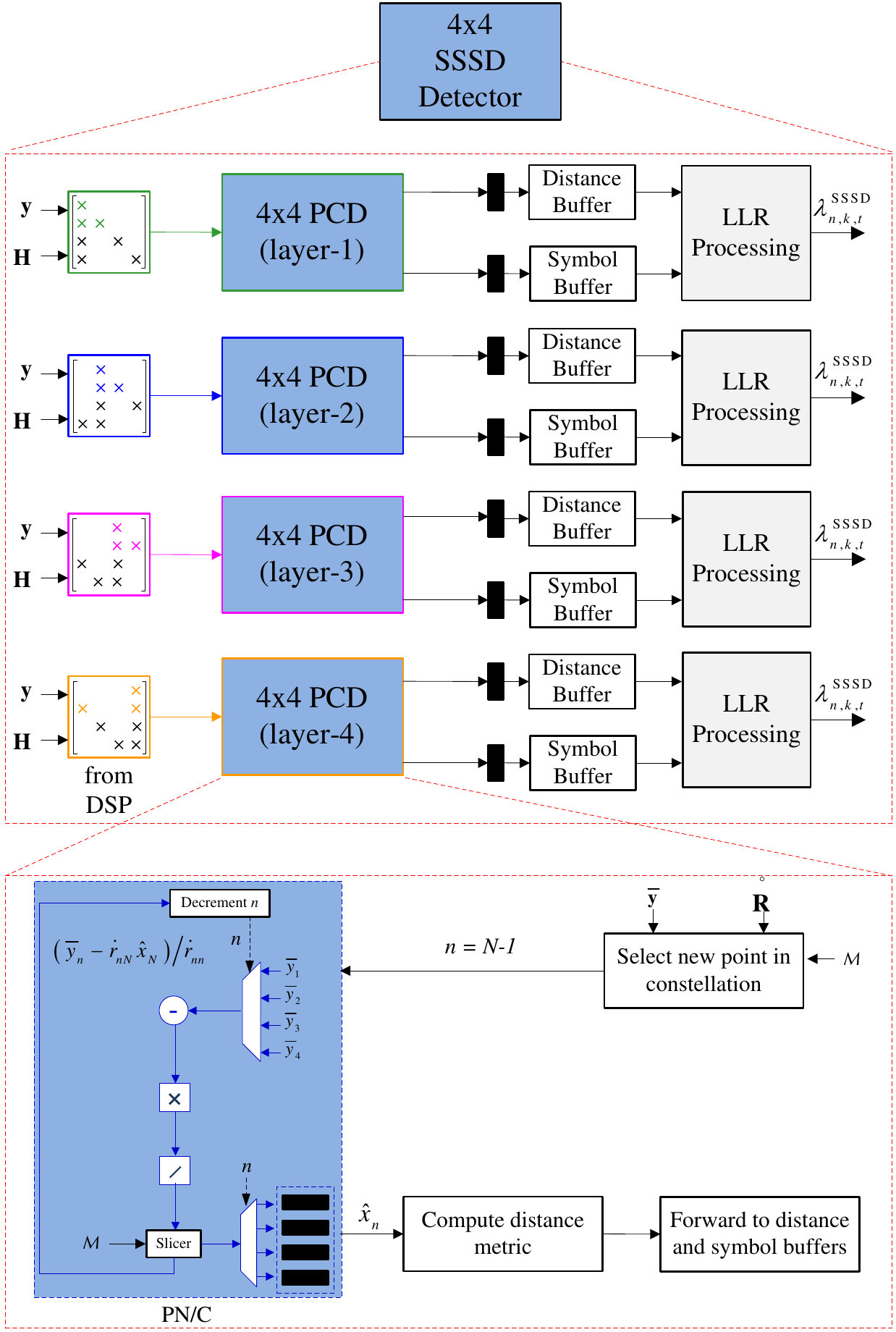}}
    \hfill
  \subfloat[]{\label{archfig:b} \includegraphics[width=0.45\linewidth]{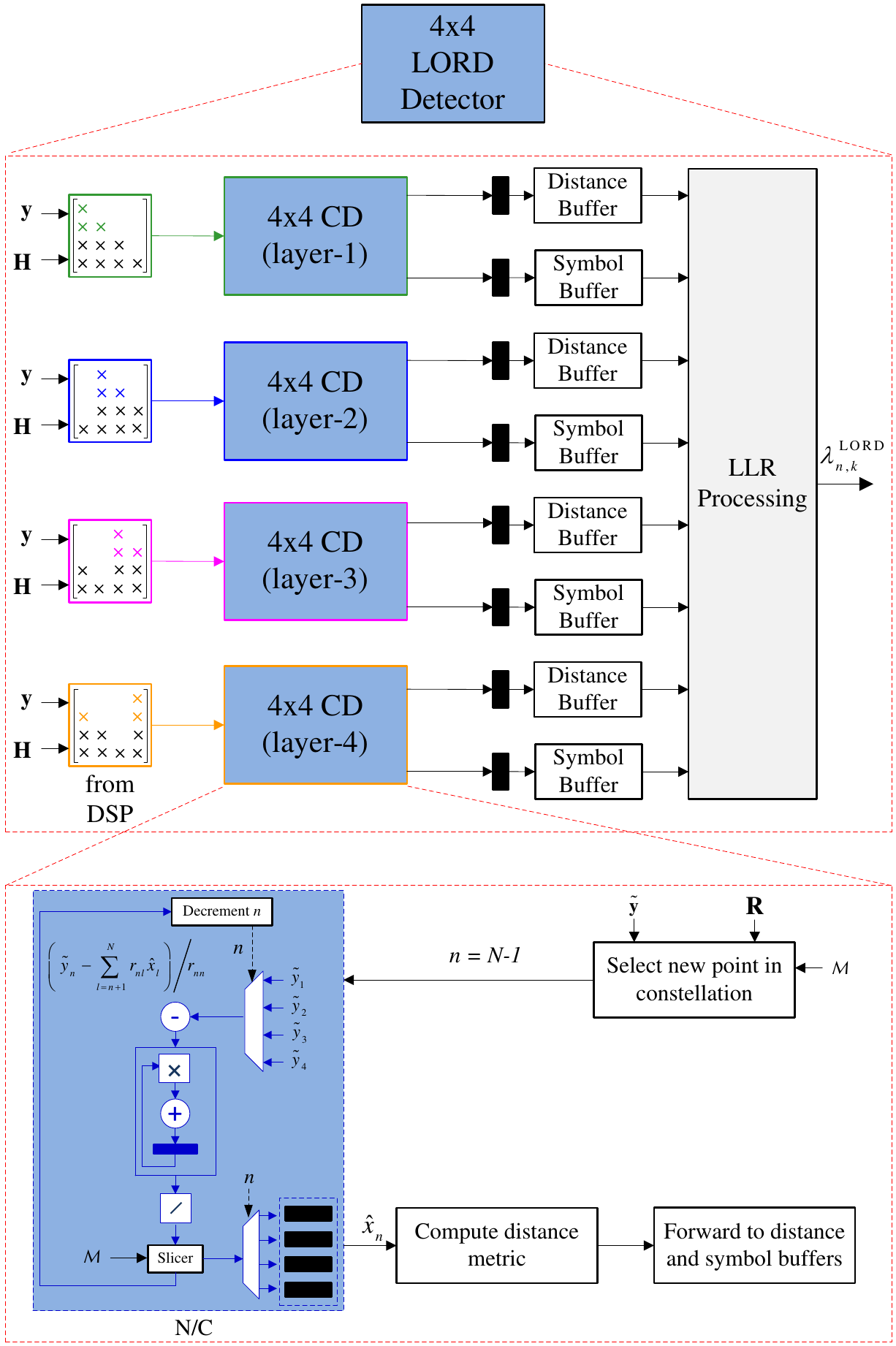}}

  \caption{Hierarchical architectural designs for (a) SO SSSD and (b) SO LORD using CDs and N/Cs as building blocks.}
  \label{archfig}
\end{figure*}

%
\section{Detector Architecture and Complexity Analysis}\label{sec:complexity}
The main motive behind channel puncturing is reducing complexity. To support this fact, a cost efficient architecture that implements the SSSD algorithm is shown in Fig.~\subref*{archfig:a}, together with a counterpart architecture that implements LORD in Fig. \subref*{archfig:b}. The architectural designs are hierarchical, showing SSSD using PCD building blocks, that themselves use PN/C, while LORD uses CD and N/C blocks.

With SSSD, the distances computed in PCD and their symbol vectors are directly forwarded to an LLR processing unit at the corresponding layer of interest. The PCD processes on all other layers can run in parallel, and the complete LLR vector will be available at the output after the processing delay of one layer. However, the LORD architecture is not fully parallelizable. Moreover, the PCD routine itself is much less complex than the CD routine because it performs fewer computations due to punctured entries.

We analyze the complexity in terms of floating-point operations (flops) based on real multiplication ($\RML$) and real addition ($\RAD$). Every time the product $\Rp\mbf{x}$ is computed, instead of $\mbf{R}\mbf{x}$, $(N-2)(N-1)/2$ complex multiplications are saved, and these saving amount to $\theta_1=(N^2-3N+2)\,\RAD+(2N^2-6N+4)\,\RML$ flops. For example, in a $16\!\times\!16$ MIMO system, $77\%$ of the multiplications required in N/C are saved in WRD-based schemes, and the savings increase to $94\%$ in a $64\!\times\!64$ MIMO system. Therefore, the SO SSD and SSSD can save $N\times\abs{\mathcal{M}}\times\theta_1$ flops compared to the SO LORD and SLORD, respectively.

\begin{table}[!t]
\centering
\footnotesize
\caption{Summary of detectors studied, and their complexity in a HO scenario} \vspace{-0.1in}
\label{table:detectors} 
\centering 
\begin{tabular}{| c || c | c | c | c |} 
\hline
Detection Scheme & QRD Cost & Puncturing Cost & Savings in Computations (flops) \\
\hline\hline
ML $\rightarrow$ PML = PCD & $\theta_2/J$ & $\theta_3/J$ & $\left(\abs{\mathcal{M}}^N-\abs{\mathcal{M}}\right)\times\theta_1$\\\hline %
N/C $\rightarrow$ PN/C & $\theta_2/J$ & $\theta_3/J$ & $\theta_1$ \\\hline %
CD $\rightarrow$ PCD & $\theta_2/J$ & $\theta_3/J$ & $\abs{\mathcal{M}}\times\theta_1$ \\\hline
LORD $\rightarrow$ SSD & $N\times\theta_2/J$ & $N\times\theta_3/J$ & $N\times(\abs{\mathcal{M}}\times\theta_1 - (4N^2+4N-2)\,\RAD-(4N^2+4N)\,\RML)$ \\\hline
SLORD $\rightarrow$ SSSD & $N\times\theta_2/J$ & $N\times\theta_3/J$ & $N\times\abs{\mathcal{M}}\times\theta_1$ \\
\hline 
\end{tabular}
\end{table}

The only computational drawback in sub-space schemes is in channel decomposition. As shown in ~\cite{2014_mansour_SPL_WLD,2014_mansour_eurasip_WLD}, regular QRD requires $\theta_2=(4N^3-N^2-N)\,\RAD+(4N^3+3N^2)\,\RML$ flops, and puncturing alone requires $\theta_3=\frac{2}{3}(8N^3-15N^2+4N-12)\,\RAD+(\frac{16}{3}N^3-7N^2+\frac{8}{3}N-20)\,\RML$ flops (this overhead was reduced in \cite{Sarieddeen_WCNC_Subspace} for SSD). However, channel matrix decompositions are only performed in the pre-processing stage of detection, and with slow fading channels, the decomposition outputs can be retained for a very large number of frames $J$.

For HO computations, PN/C saves $\theta_1$ flops compared to N/C, PCD saves $\abs{\mathcal{M}}\times\theta_1$ flops compared to CD, SSSD saves $N\times\abs{\mathcal{M}}\times\theta_1$ flops compared to SLORD, and SSD saves $N\times(\abs{\mathcal{M}}\times\theta_1 - (4N^2+4N-2)\,\RAD-(4N^2+4N)\,\RML)$ flops compared to LORD, where the subtracted terms in the latter account for distance computations in~\eqref{eq:SSDout}. These results are summarized in Table \ref{table:detectors}. Note that the substantial savings with PML are based on the fact that PML and PCD are identical. PML has no practical significance, and it is only included as a reference. Furthermore, extra memory is needed with LORD to compute the global minimum distance for every bit after layer processing \cite{2015_mansour_JSP_2x2QAM}, where it is required to store not only distances, but also their corresponding symbol vectors. The WRD-based approaches are thus computationally efficient, especially with slow fading channels and high order modulation constellations.

\begin{figure*}[t]
  \centering
  \subfloat[$4\times4$ MIMO]{\label{hardfig:a} \includegraphics[width=0.48\linewidth]{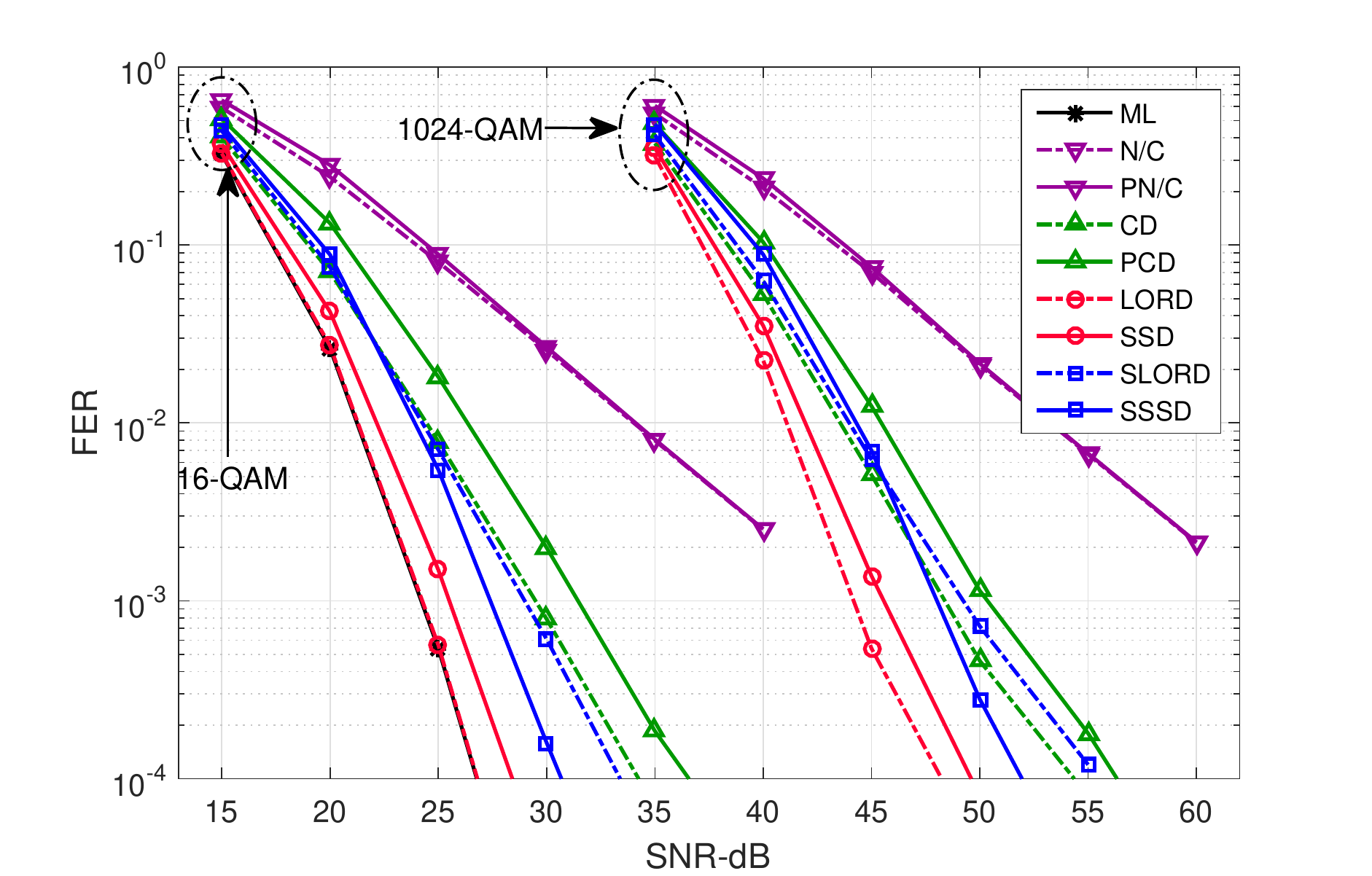}}
  \hfill
  \subfloat[$8\times8$ MIMO]{\label{hardfig:b} \includegraphics[width=0.48\linewidth]{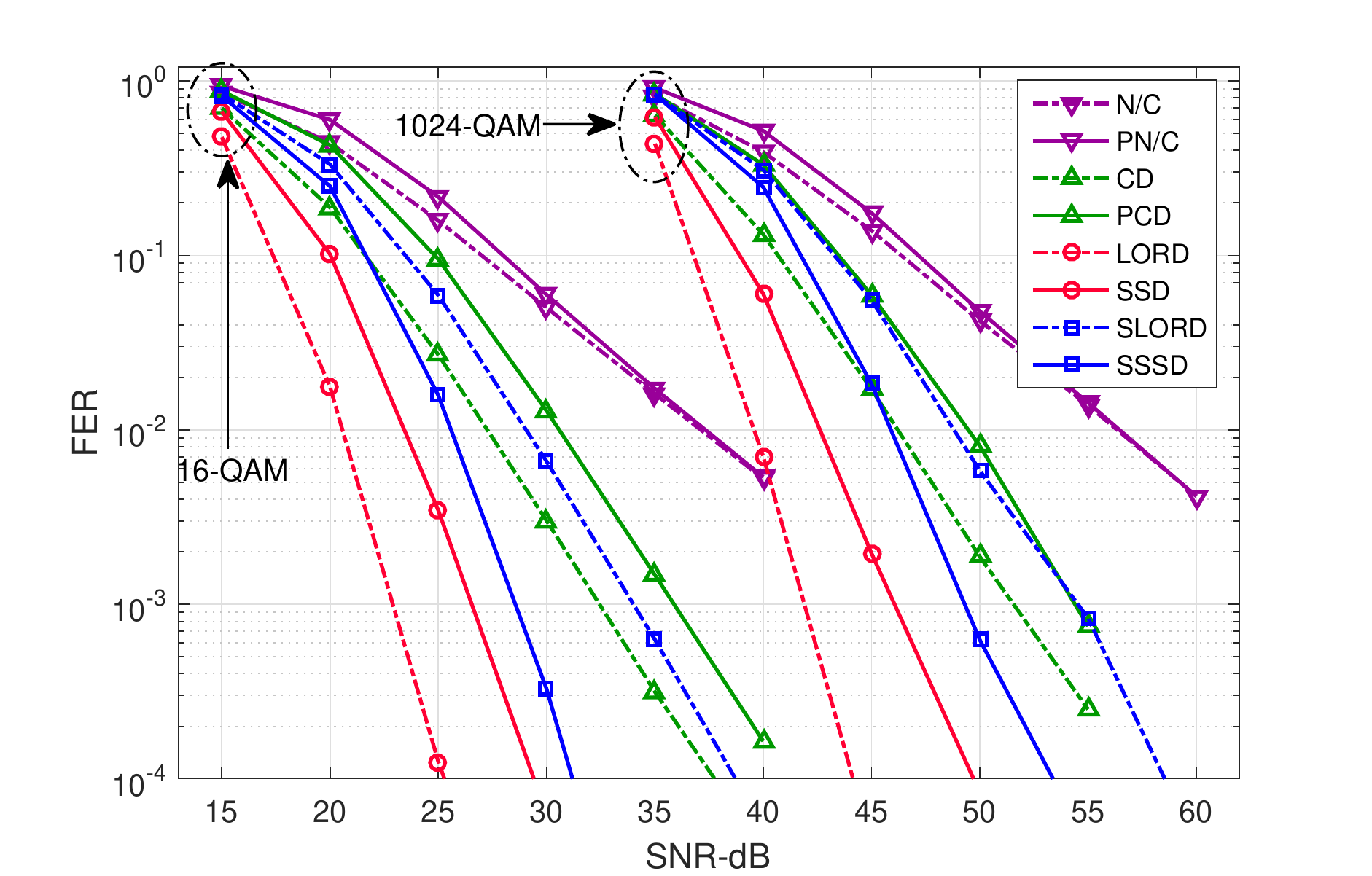}}

  \subfloat[$16\times16$ MIMO]{\label{hardfig:c} \includegraphics[width=0.48\linewidth]{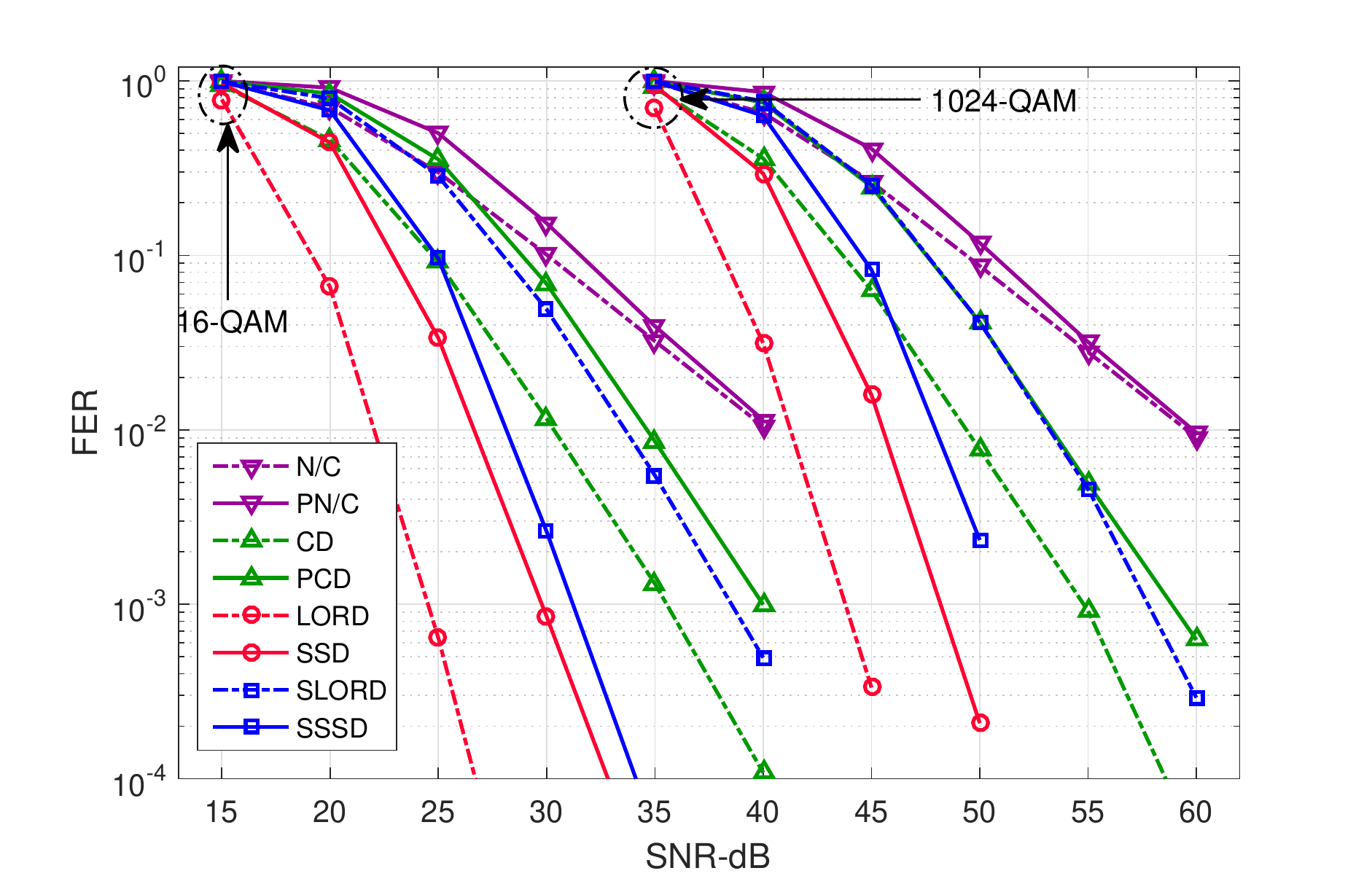}}
  \hfill
  \subfloat[$64\times64$ MIMO]{\label{hardfig:d} \includegraphics[width=0.48\linewidth]{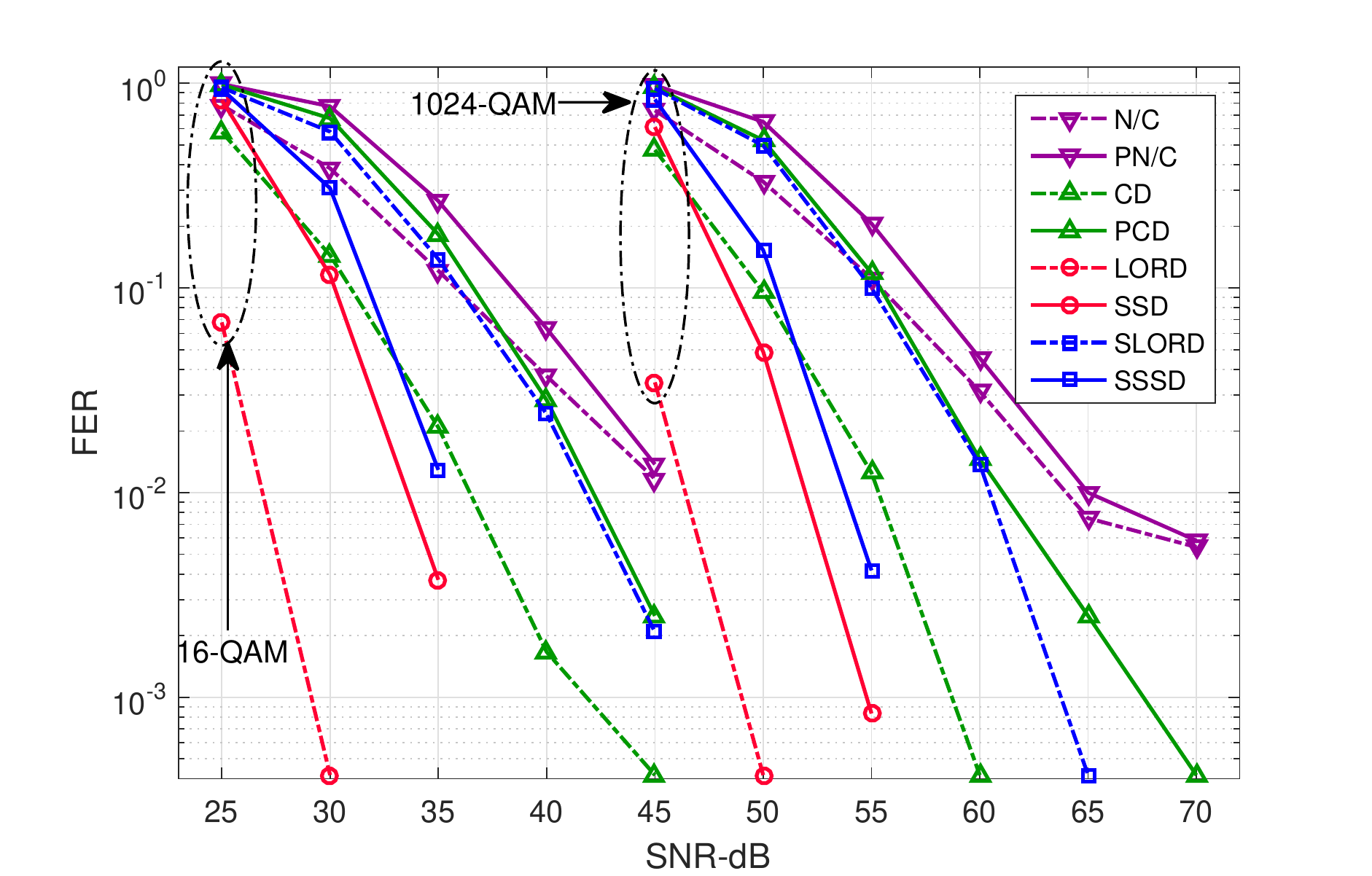}}

  \caption{HO FER performance of the studied detectors with $16$-QAM and $1024$-QAM.}
  \label{f:hard}
\end{figure*}

%
\section{Simulation Results}\label{sec:results}
The detection schemes considered above were simulated following the system model of Sec.~\ref{sec:sysmodel}. Both HO and SO scenarios were considered, where in the latter turbo coding was used, with a code rate of $1/2$ and $8$ decoding iterations. In addition to $\mbf{H}$, which corresponds to rich scattering, we considered another channel $\mbf{H}_{c}$, that accounts for antenna correlation. The effective channel matrices are related by $\mbf{H}_{c}\!=\!\mbf{R}_{r}^{1/2}\mbf{H}\mbf{R}_{t}^{1/2}$, where $\mbf{R}_{t}$ and $\mbf{R}_{r}$ are the transmit and receive antenna correlation matrices, respectively, with correlation factors $\alpha=\beta=0.9$. We assume, for convenience, the generic exponential model~\cite{loyka2001channel}. Hence, in the case of $4\!\times\!4$ MIMO, we have,
\begin{equation}\label{matrix}
 \mbf{R}_{t} = \begin{bmatrix}
    1 & \alpha & \alpha^2 & \alpha^3  \\
    \alpha & 1 & \alpha & \alpha^2  \\
    \alpha^2 & \alpha & 1 & \alpha \\
    \alpha^3 & \alpha^2 & \alpha & 1
\end{bmatrix},\ \ \ \  \mbf{R}_{r} = \begin{bmatrix}
    1 & \beta & \beta^2 & \beta^3  \\
    \beta & 1 & \beta & \beta^2  \\
    \beta^2 & \beta & 1 & \beta \\
    \beta^3 & \beta^2 & \beta & 1
\end{bmatrix}.
\end{equation}

\begin{figure*}[t]
  \centering
  \subfloat[$8\times8$ MIMO]{\label{softfig:a} \includegraphics[width=0.48\linewidth]{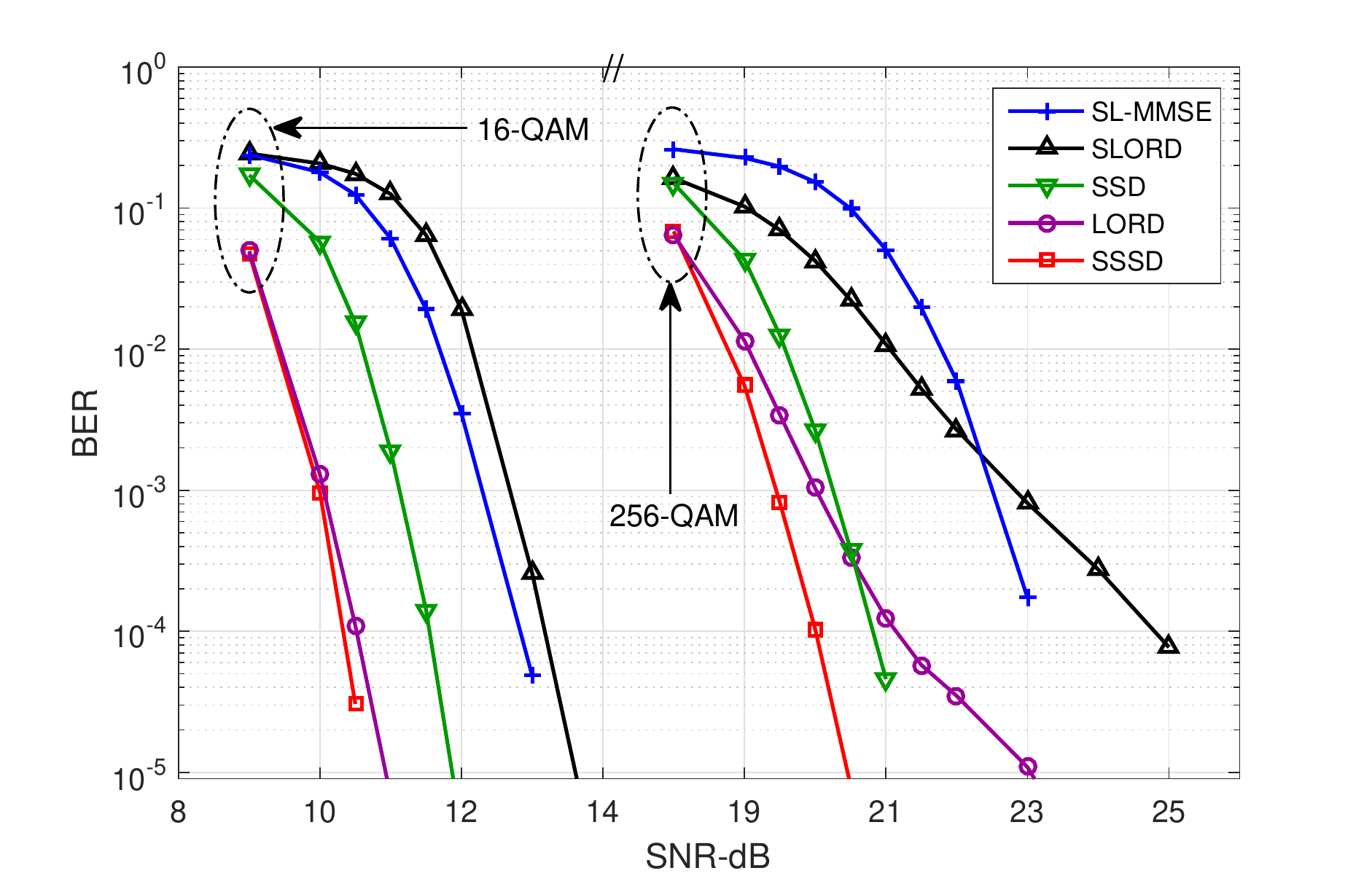}}
  \hfill
  \subfloat[$16\times16$ MIMO]{\label{softfig:b} \includegraphics[width=0.48\linewidth]{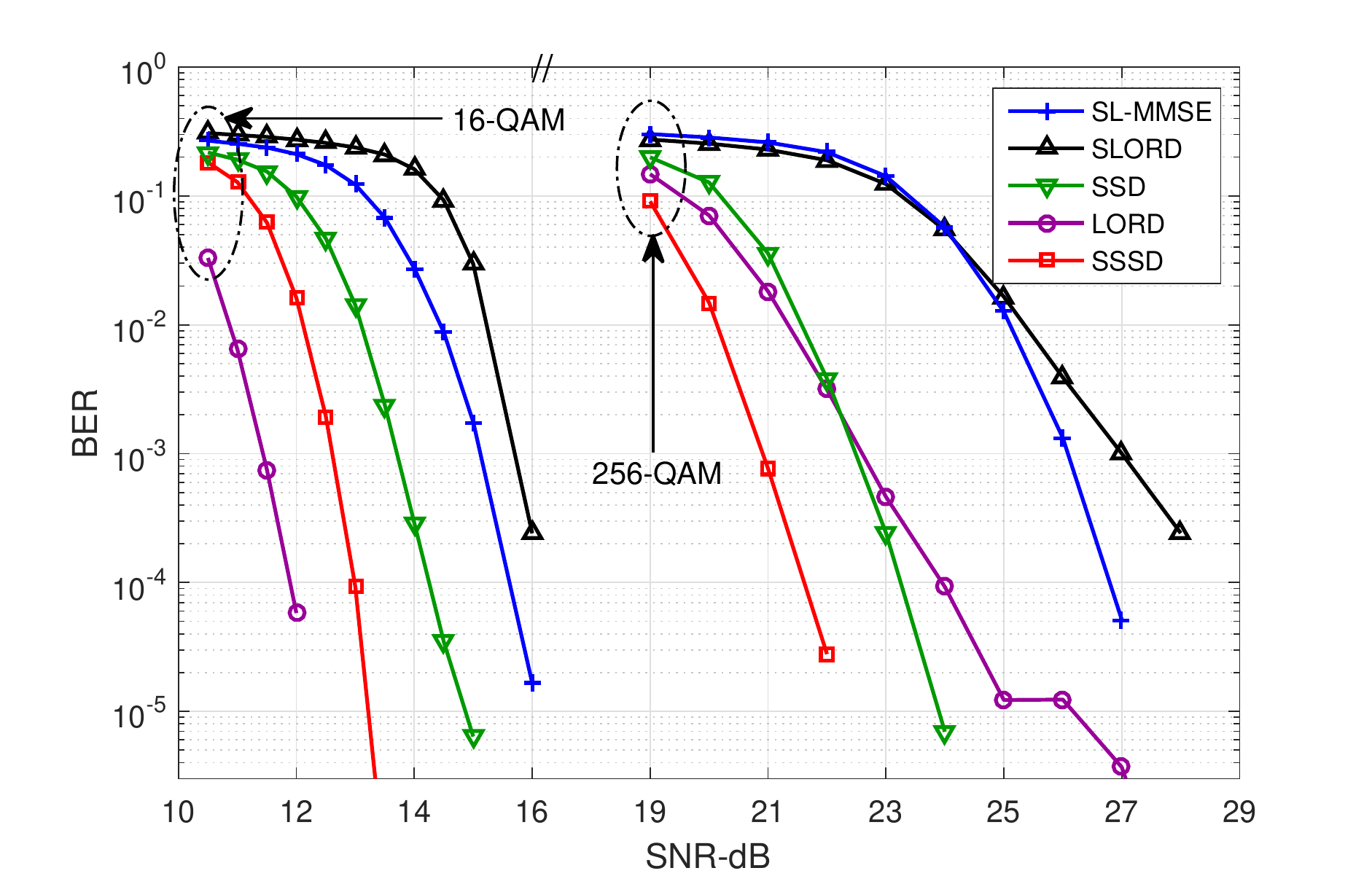}}

  \subfloat[$64\times64$ MIMO]{\label{softfig:c} \includegraphics[width=0.48\linewidth]{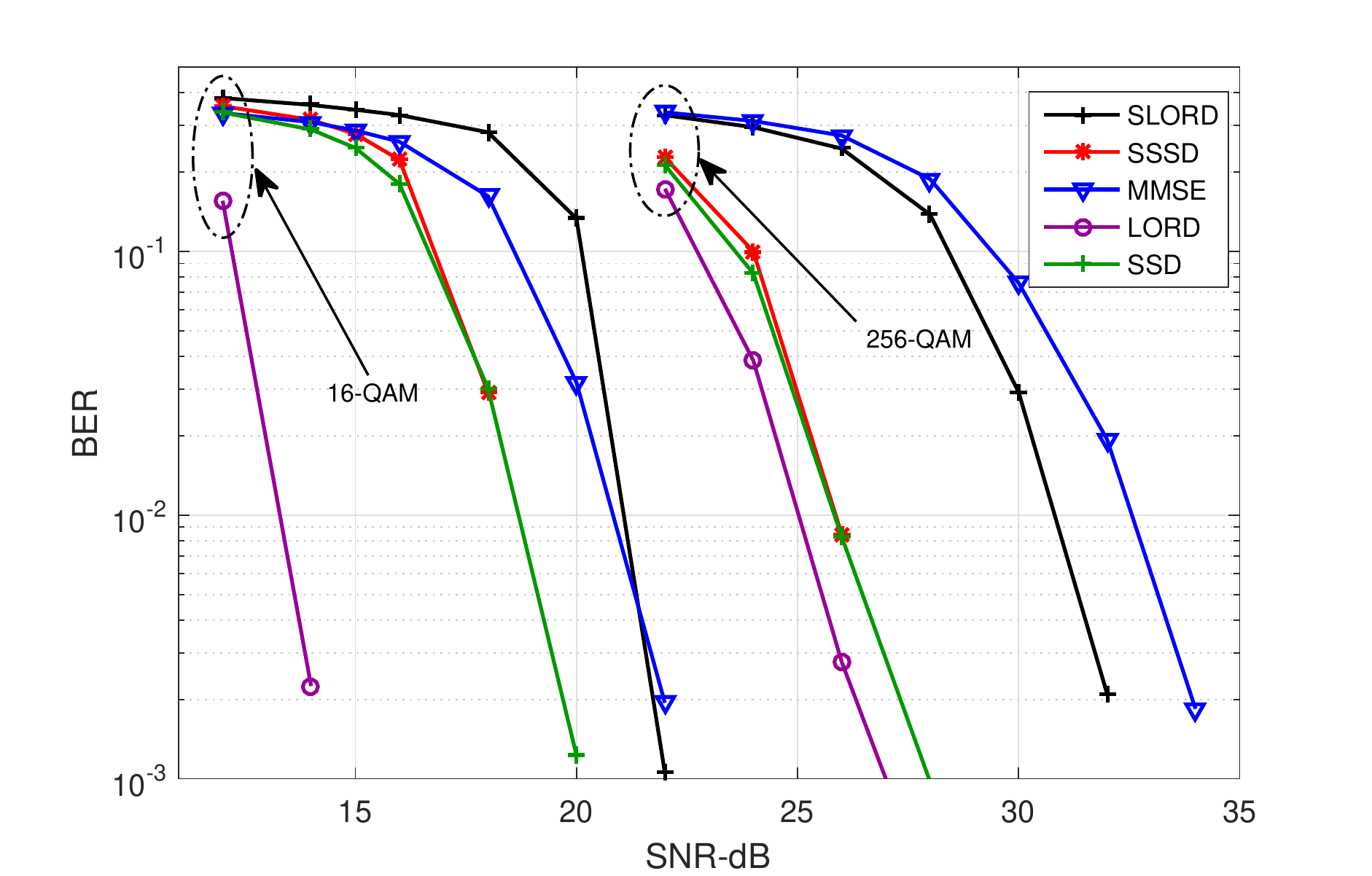}}
  \hfill
  \subfloat[$128\times128$ MIMO]{\label{softfig:d} \includegraphics[width=0.48\linewidth]{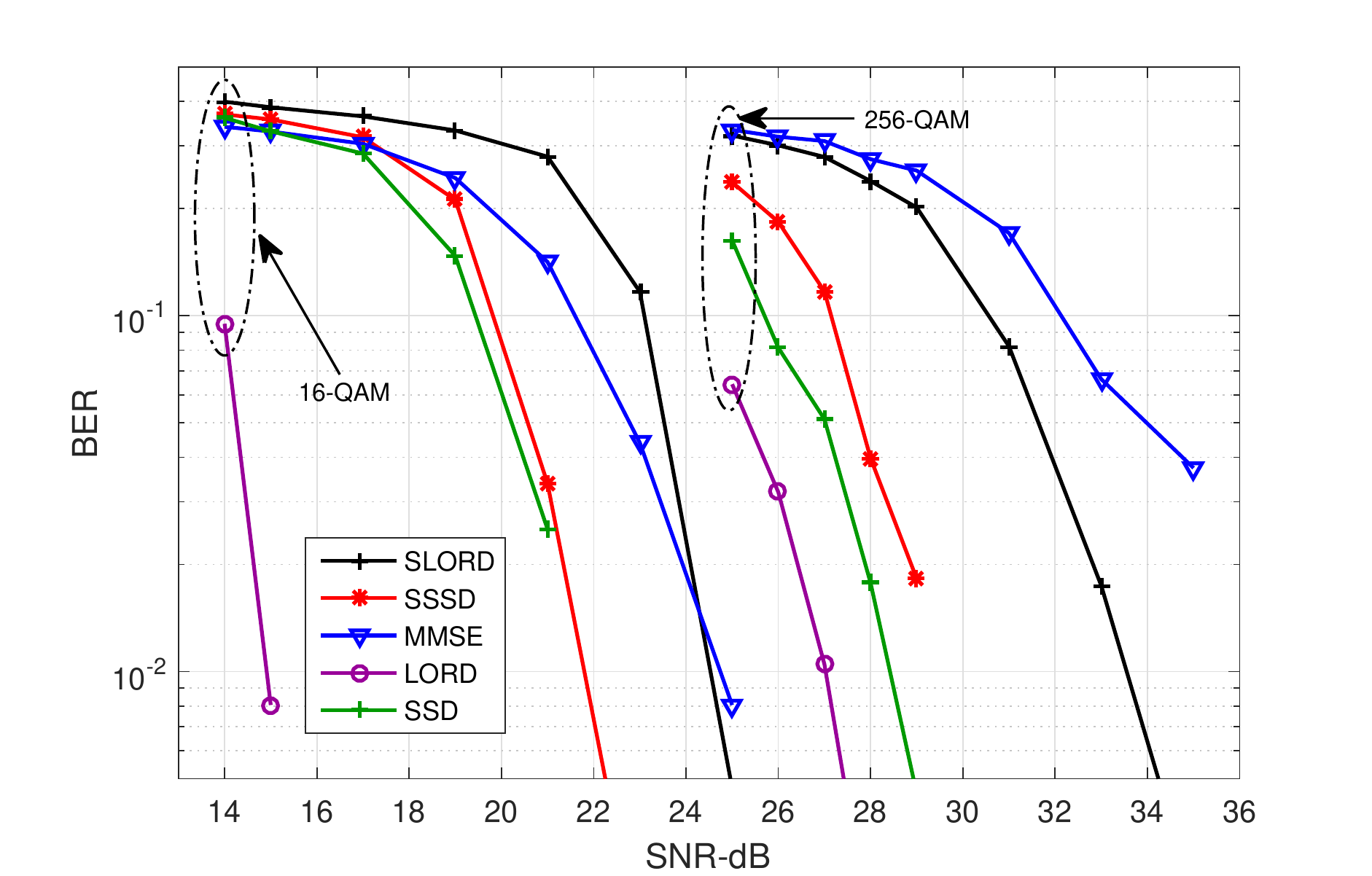}}

  \caption{SO BER performance of the studied detectors with $16$-QAM and $256$-QAM.}
  \label{f:soft}
\end{figure*}

\begin{figure*}[t]
  \centering
  \subfloat[$4\times4$ MIMO]{\label{corrsoftfig:a} \includegraphics[width=0.48\linewidth]{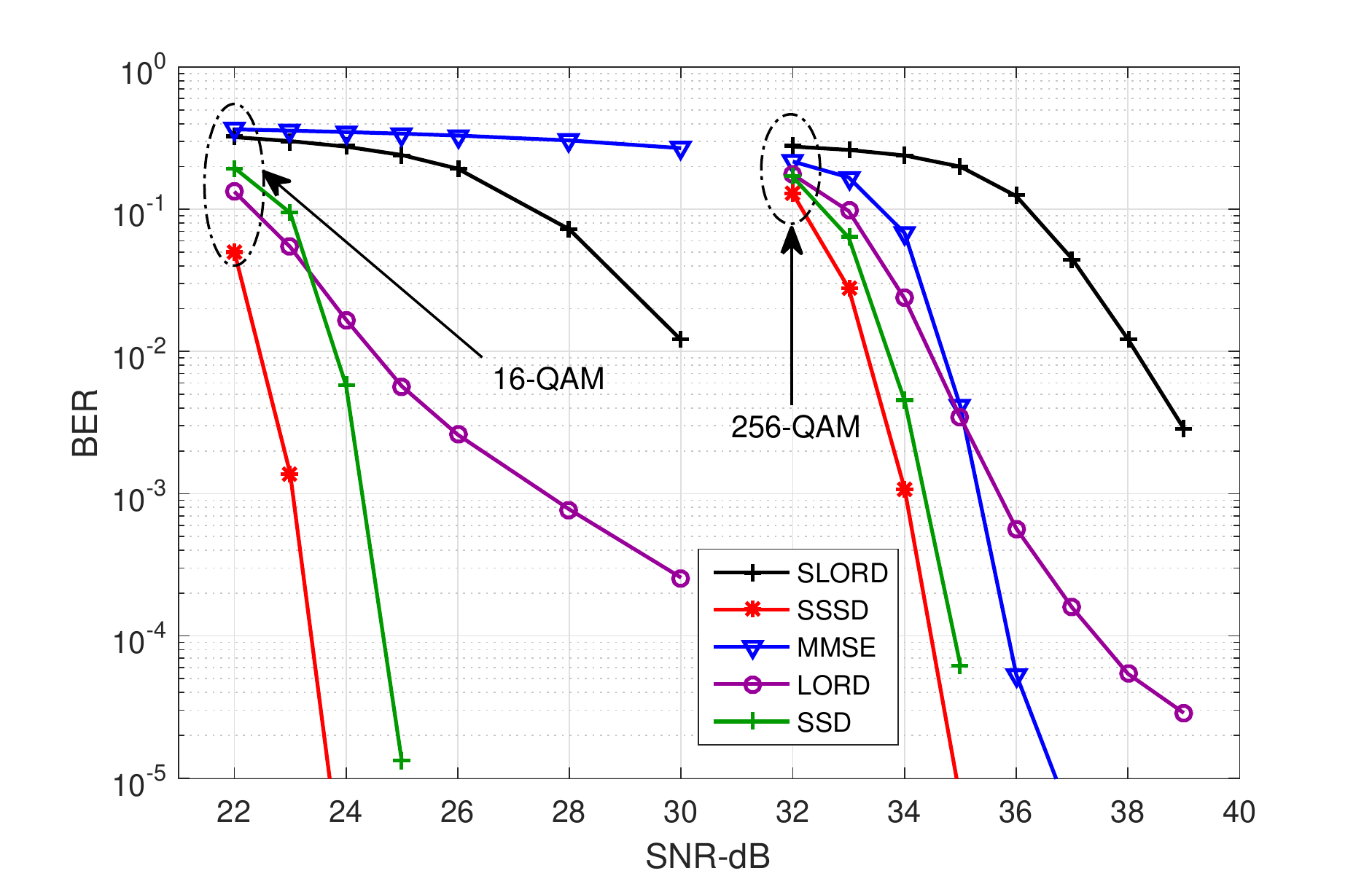}}
  \hfill
  \subfloat[$8\times8$ MIMO]{\label{corrsoftfig:b} \includegraphics[width=0.48\linewidth]{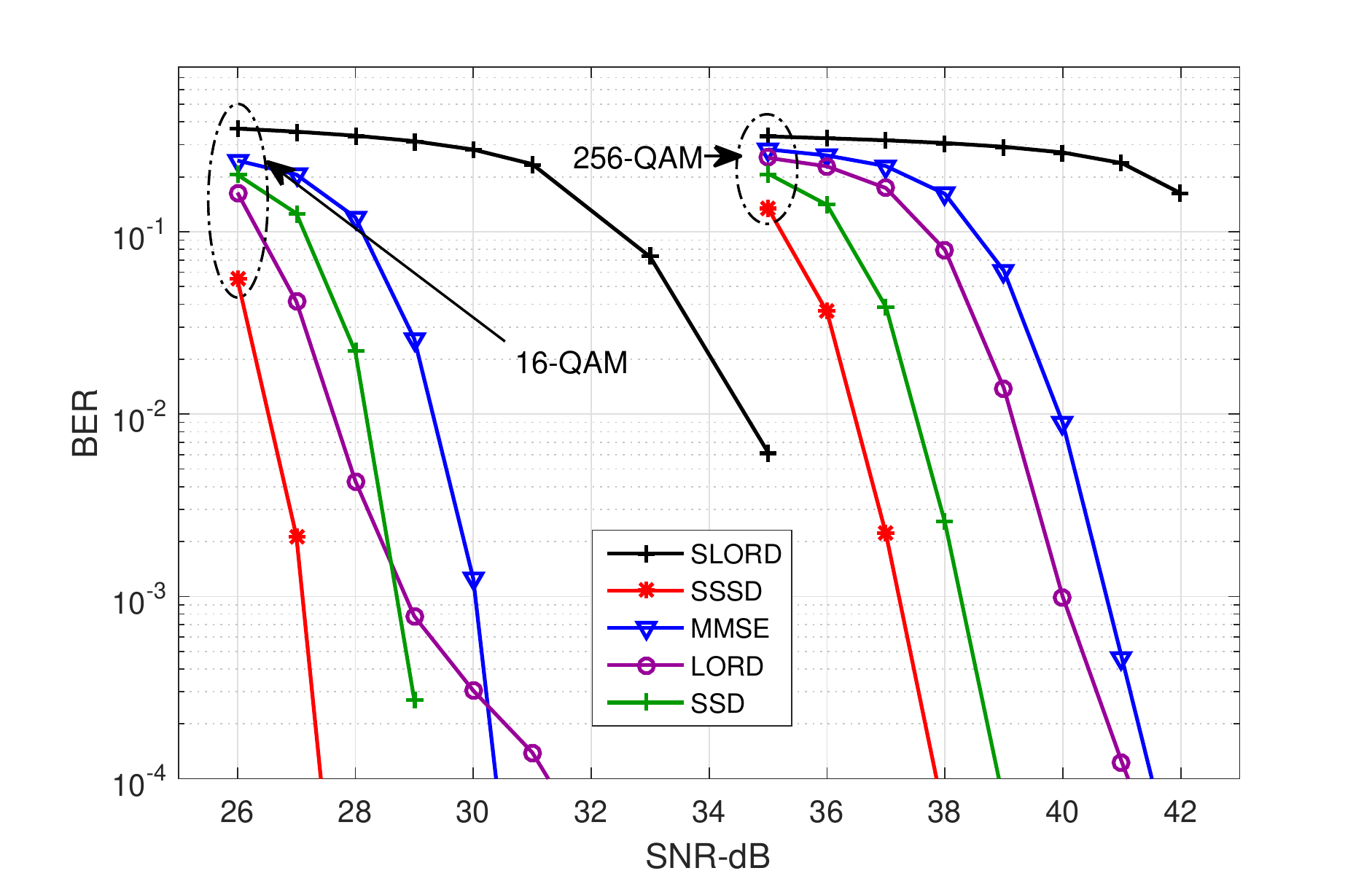}}

  \subfloat[$16\times16$ MIMO]{\label{corrsoftfig:c} \includegraphics[width=0.48\linewidth]{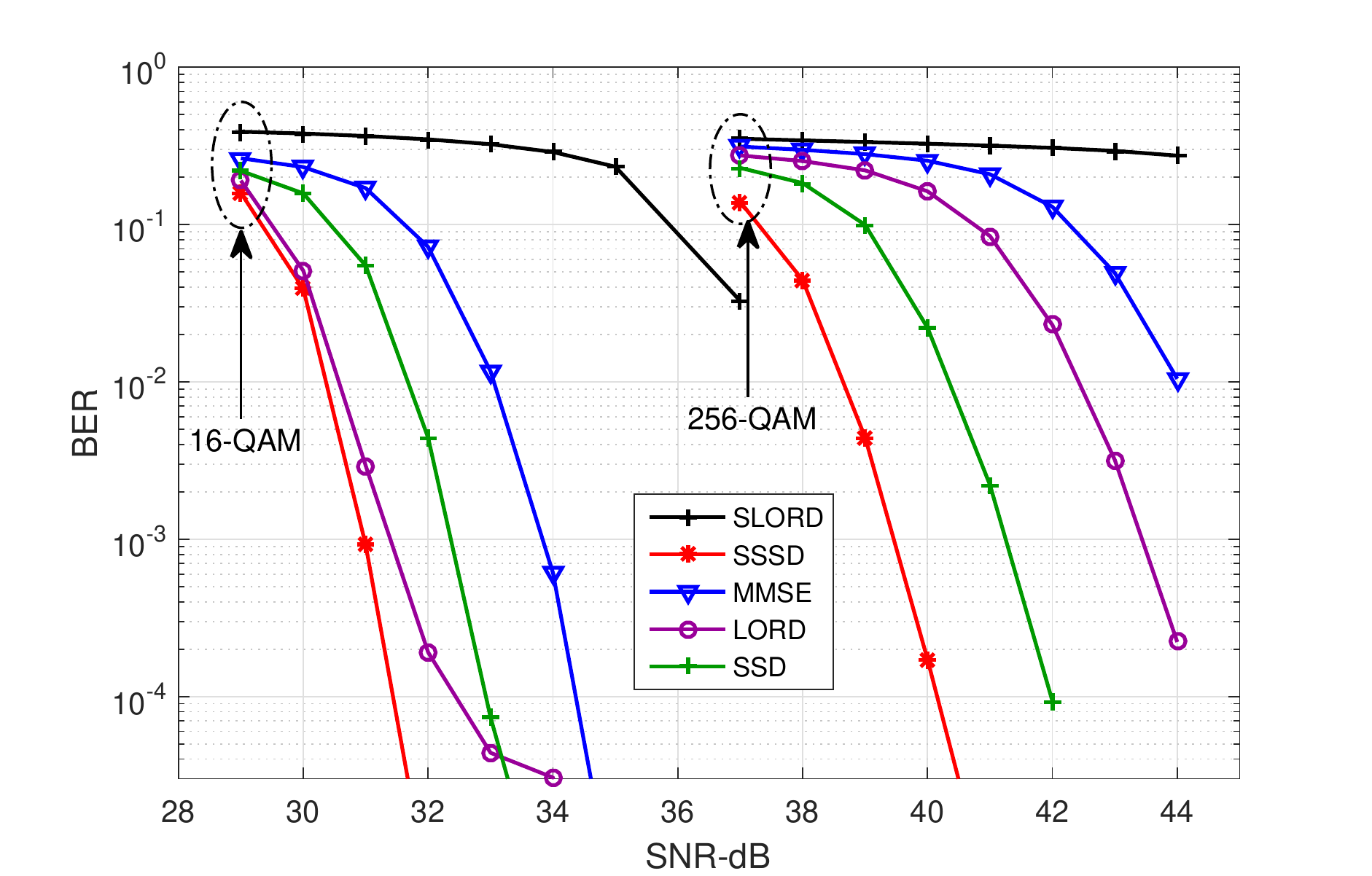}}
  \hfill
  \subfloat[$64\times64$ MIMO]{\label{corrsoftfig:d} \includegraphics[width=0.48\linewidth]{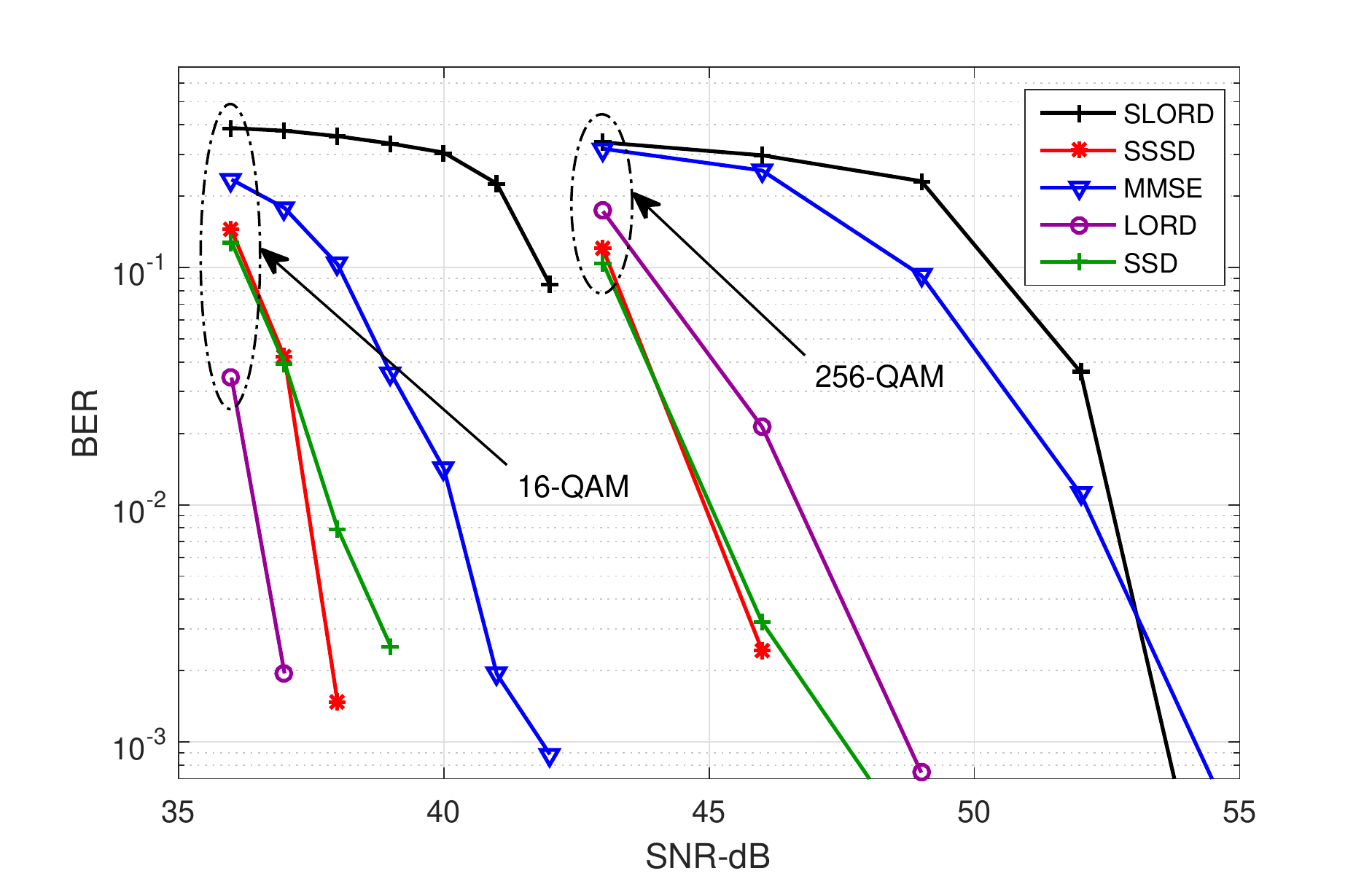}}

  \caption{SO BER performance of the studied detectors under channel correlation with $16$-QAM and $256$-QAM.}
  \label{f:corrsoft}
\end{figure*}

Fig.~\ref{f:hard} shows the HO frame error rate (FER) performance with various MIMO configurations, when $\mathcal{M}$ is $16$-QAM and $1024$-QAM \cite{Sarieddeen_GlobalSIP_1024}. In the context of $4\!\times\!4$ MIMO (Fig. \subref*{hardfig:a}), the performance degradation in PN/C compared to N/C is negligible, the chase detectors cut the gap between N/C and ML in half, and the PCD introduces a $\unit[2]{dB}$ loss compared to the CD. Moreover, while LORD achieves exact ML performance, SSD lags behind by also $\unit[2]{dB}$. The relative performances of the detectors are maintained with very large constellations. Figures \subref*{hardfig:b}, \subref*{hardfig:c}, and \subref*{hardfig:d} then show the performance of the proposed schemes in the context of $8\!\times\!8$ MIMO, $16\!\times\!16$ MIMO, and $64\!\times\!64$ MIMO, respectively. The relative performances are maintained, but the gap between WRD and QRD-based schemes increases from $\unit[2]{dB}$, to $\unit[4]{dB}$, $\unit[5]{dB}$, and $\unit[7]{dB}$, respectively. The SSSD was the only WRD-based detector to achieve a performance gain, compared to SLORD.

Fig.~\ref{f:soft} shows the BER performance of the studied SO detectors, compared to a reference low-complexity MMSE detector \cite{Studer_MMSE}, with various MIMO configurations, when $\mathcal{M}$ is $16$-QAM and $256$-QAM. With relatively low MIMO orders, the SSSD (or SO PCD) outperforms LORD, and so does the SSD with high order modulation types, while SLORD and MMSE lag behind. For example, the SO PCD achieves a $\unit[2.5]{dB}$ gain compared to LORD, at a BER of $10^{-4}$ in $16\!\times\!16$ MIMO with $256$-QAM. However, with high order MIMO, SSSD can not beat LORD ($\unit[5]{dB}$ and $\unit[7]{dB}$ gaps are noticed with $16$-QAM). Nevertheless, at very high MIMO orders, the reduction in complexity with WRD-based detectors is particularly large, and the gap in performance can be as low as $\unit[1]{dB}$ or $\unit[2]{dB}$ with $256$-QAM, where the effect of interference is reduced with high order constellations (\ref{sec:diversitySSSD}).

Fig.~\ref{f:corrsoft} shows the SO BER performance of the detectors under high channel correlation. Sub-space detectors, SSD and SSSD, outperform the much more complex LORD. It is only at very high MIMO orders with low order modulation types that LORD slightly outperforms SSSD. This declares the SSSD the winning detector in the presence of channel correlation.

%
\section{Conclusions}\label{sec:conclusion}
A family of low-complexity MIMO detectors that employ punctured QRD in lieu of regular QRD has been proposed and studied both analytically, by deriving bounds on the achievable diversity gains and error probability, and empirically through simulations. The proposed detectors have been shown to achieve significant computational savings in the context of large MIMO systems, while at the same time achieving the same diversity gains as their QRD-based counterparts. Furthermore, significant performance gains have been observed with highly correlated channels. An architectural design has been proposed, by using the detectors of lower complexity as building blocks in their more complex extensions, and it has been established that the proposed schemes scale up efficiently both in the number of antennas and the constellation size. In particular, soft-output per-layer sub-space detection has been shown to achieve a $\unit[2.5]{dB}$ $\mathsf{SNR}$ gain in $256$-QAM $16\!\times\!16$ MIMO, while saving $77\%$ of nulling-and-cancellation computations.

It can be argued that the proposed schemes in this paper are better candidates for large MIMO detection than reference detectors in the literature when complexity is taken into consideration. For example, detection based on local search \cite{Vardhan_2008,Li_2010} does not achieve near-ML diversity, nor does it scale up efficiently with high order modulation constellations. Similarly, heuristic tabu search algorithms \cite{Datta_2010,Srinidhi_2011} do not perform well with high order modulation types, and their performance is hard to track analytically. Detectors based on message passing on graphical models \cite{Goldberger_2011,Narasimhan_2014} have been recently extended to support high order modulation types; however, they are better suited for joint iterative detection and decoding schemes. Furthermore, with lattice reduction~\cite{Wubben_2011,Zhou_LR_2013}, processing large channel matrices using conventional reduction schemes is costly, while low complexity schemes such as element-based lattice reduction incur performance degradation. Moreover, computing SO LLRs is not straightforward. Finally, detection using Monte Carlo sampling \cite{Datta_2013} is clearly outperformed by the proposed schemes.

 \appendices

  \section{BER of L-QAM Over $d$-Fold Diversity}
  \label{FirstAppendix}

The function $G(d,\gamma,L)$ provides the average BER of L-QAM over a $d$-fold diversity Rayleigh fading channel with mean branch SNR $\gamma$ \cite{Kim_1997,Kim_2008}:

\begin{equation}\label{eq:app1_1}
    G(d,\gamma,L) = \frac{\sqrt{L}-1}{L} \left[ (\sqrt{L}-1) + 4I_1 - (\sqrt{L}-1)I_2 \right]
\end{equation}
\begin{equation}\label{eq:app1_2}
    I_1 = \left[ \frac{1}{2} \left( 1-\mu\right)\right]^d \sum_{k=0}^{d-1} \left( \begin{array}{c} d-1+k \\ k \end{array} \right)\left[ \frac{1}{2} \left( 1+\mu\right) \right]^k
\end{equation}
\begin{equation}\label{eq:app1_3}
\mu=\sqrt{\frac{\beta\gamma}{1+\beta\gamma}}, \ \ \ \ \ \ \beta = \frac{3\log_2(L)}{2(L-1)}
\end{equation}
\[
  I_2=\begin{cases}

                \displaystyle  \ \frac{4}{\pi} \ \mu \tan^{-1}\mu  \ \ \ \ \text{for} \ \ d = 1 \\~\\
                \vspace{+0.05in}
                \displaystyle  \ \frac{4}{\pi}  \sum_{k=0}^{d-1} \frac{(2k)!}{2^{2k}(k!)^2} \bigg[ \left(\frac{1}{1+\beta\gamma}\right)^k \mu \tan^{-1}\mu \bigg] \\ \vspace{+0.05in}
                \displaystyle \ + \frac{2}{\pi}  \sum_{k=1}^{d-1} \frac{(2k)!}{2^{2k}(k!)^2} \bigg[ \sum_{v=1}^{k} \frac{2^{2v}v!}{(2v)!} \left(\frac{1}{1+\beta\gamma}\right)^{k-v+1} \\
                \displaystyle \ \times \left(\frac{\beta\gamma}{1+2\beta\gamma}\right) (v-1)! \left(\frac{1}{1+2\beta\gamma}\right)^{v-1} \bigg]  \ \ \ \ \text{for} \ \ d \geq 2

            \end{cases}
\]

Note that when using the function $G(d,\gamma,L)$ in equations \eqref{eq:berSIC} and \eqref{eq:berPSIC}, the constant term 4, which represents the variance of the error that is caused by wrong decisions on lower layers, should be modified. With L-QAM, and assuming normalized constellations, an error at a lower layer will more likely incur a noise of variance $(2/\log_2(L))^2$ at upper layers.


\ifCLASSOPTIONcaptionsoff
  \newpage
\fi

%

\end{document}